\documentclass[english,twocolumn,reprint,aip,jcp]{revtex4-1}
\usepackage[latin9]{inputenc}
\usepackage{amsmath}
\usepackage{amssymb}
\usepackage{graphicx}
\usepackage{wasysym}
\usepackage{babel}
\begin{document}

\title{Inchworm Monte Carlo for exact non-adiabatic dynamics II. Benchmarks
and comparison with established methods}

\author{Hsing-Ta Chen}

\affiliation{Department of Chemistry, Columbia University, New York, New York
10027, U.S.A.}

\affiliation{The Raymond and Beverly Sackler Center for Computational Molecular
and Materials Science, Tel Aviv University, Tel Aviv 69978, Israel}

\author{Guy Cohen}

\affiliation{The Raymond and Beverly Sackler Center for Computational Molecular
and Materials Science, Tel Aviv University, Tel Aviv 69978, Israel}

\affiliation{School of Chemistry, The Sackler Faculty of Exact Sciences, Tel Aviv
University, Tel Aviv 69978, Israel}

\author{David R. Reichman}

\affiliation{Department of Chemistry, Columbia University, New York, New York
10027, U.S.A.}
\begin{abstract}
In this second paper of a two part series, we present extensive benchmark
results for two different inchworm Monte Carlo expansions for the
spin\textendash boson model. Our results are compared to previously
developed numerically exact approaches for this problem. A detailed
discussion of convergence and error propagation is presented. Our
results and analysis allow for an understanding of the benefits and
drawbacks of inchworm Monte Carlo compared to other approaches for
exact real-time non-adiabatic quantum dynamics.
\end{abstract}
\maketitle

\section{Introduction\label{sec:Introduction}}

The spin\textendash boson model is perhaps the most basic example
of a quantum dissipative system.\cite{Leggett1987Dynamicsdissipativetwo}
Despite its simplicity, the model endures for several reasons. First,
the spin\textendash boson rather faithfully represents the spectrum
of realistic behaviors associated with the relaxation of a small quantum
system connected to a heat bath.\cite{Weiss1999Quantumdissipativesystems}
While the model employs seemingly unrealistic features such as linear
coupling to a harmonic reservoir, even anharmonic systems may be mapped
to this form of environmental interaction within linear response theory.\cite{Bader1990Rolenucleartunneling,Warshel1989Dispersedpolaronsimulations,Warshel1986Simulationdynamicselectron,Makri1999LinearResponseApproximation}
This generality explains the wide usage of the spin\textendash boson
paradigm in the modeling of systems ranging from charge and energy
transfer in condensed phases and biological systems\cite{Garg1985Effectfrictionelectron,Georgievskii1999Linearresponsein,Adolphs2006HowProteinsTrigger,Engel2007Evidencewavelikeenergy,Lee2007Coherencedynamicsin,Panitchayangkoon2010Longlivedquantum,Collini2009CoherentIntrachainEnergy,Bredas2009ExcitonsSurfAlong}
to the relaxation of dilute impurities in the solid state\cite{Fisher1985QuantumBrownianmotion,ProkofEv1996Quantumrelaxationmagnetisation,ProkofEv1998Lowtemperaturequantum}
and in Josephson junction arrays.\cite{Goldstein2013InelasticMicrowavePhoton}
In the rotating wave approximation, the spin\textendash boson model
is reduced to the Jaynes\textendash Cummings model, which is of great
importance in quantum optics.\cite{Fink2008ClimbingJaynesCummingsladder,Solano2003StrongDrivingAssisted,Casanova2010DeepStrongCoupling}

A second major reason for the continued study of the spin\textendash boson
model resides in the fact that it cannot be solved analytically and
presents numerical challenges for exact quantum dynamics methodologies.
Thus the model has become a canonical benchmark for both approximate
and exact dynamical approaches. After approximately two decades of
research, several numerically exact approaches have emerged which
are capable of long-time simulation of time-dependent observables
of the spin\textendash boson Hamiltonian, at least for many important
regimes of the model. These methods differ from one another with respect
to their formulations, their scaling properties, their complexity
and their generality. In this paper we will concern ourselves with
benchmarks from three commonly used approaches: the quasi-adiabatic
path integral (QUAPI) approach,\cite{Makarov1994Pathintegralsdissipative,Makri1995Numericalpathintegral,Makri1995Tensorpropagatoriterative,Makri1996Longtimequantum}
the hierarchical equations-of-motion (HEOM) method,\cite{Tanimura1989TimeEvolutionQuantum,Ishizaki2009Unifiedtreatmentquantum,Struempfer2012OpenQuantumDynamics}
and the multi-layer, multi-configuration time-dependent Hartree (ML-MCTDH)
approach.\cite{Thoss2001Selfconsistenthybrid,Wang2001Systematicconvergencein,Wang2003Multilayerformulationmulticonfiguration}
There are fewer calculations of spin\textendash boson dynamics done
by stationary-phase Monte Carlo,\cite{Mak1990Solvingsignproblem,Mak1991Coherentincoherenttransition,Egger1992QuantumMonteCarlo,Egger1994Lowtemperaturedynamical,Egger2000PathintegralMonte}
though this technique also yields numerically exact results.

We will focus on diagrammatic Monte Carlo methods,\cite{Prokofev1998PolaronProblemby,Prokofev2008BolddiagrammaticMonte,Prokofev2008Fermipolaronproblem,VanHoucke2010DiagrammaticMonteCarlo}
a dominant and widely applicable formalism for addressing impurity
models. These methods allow for facile computation of equilibrium
observables in general impurity-type problems,\cite{Gull2011ContinuoustimeMonte}
including those unrelated to the spin\textendash boson model. They
have also been used for non-equilibrium problems,\cite{Muehlbacher2008Realtimepath,Werner2009DiagrammaticMonteCarlo,Cohen2014Greensfunctionsfroma,Cohen2014Greensfunctionsfrom,Cohen2013Numericallyexactlong}
where they are limited in applicability by the dynamical sign problem.
Recently, we have introduced an exact, real-time diagrammatic Monte
Carlo approach called the inchworm algorithm\cite{Cohen2015Tamingdynamicalsign}
which greatly suppresses the dynamical sign problem. In the companion
paper to the work presented here, we have elaborated upon and expanded
the scope of the approach, using the spin\textendash boson model as
a concrete example.

In this work we compare the results of the inchworm algorithm to those
produced by the other methodologies mentioned above in essentially
all regimes of interest. Our results allow us to compare and contrast
the strengths and weaknesses of the relative approaches. We demonstrate
that the inchworm algorithm is competitive with the most advanced
real-time approaches and is capable of producing converged long-time
results even in some regimes difficult for several prominent approaches.
The success of the inchworm algorithm as outlined in this work paves
the way for a host of novel applications, a few of which we enumerate
at the conclusion of this paper.

The paper is organized as follows. In Sec.~\ref{sec:Model-and-method},
we specify details of the spin\textendash boson model and provide
an analysis of convergence for the system\textendash bath coupling
inchworm (SBCI) and the diabatic coupling cumulant inchworm (DCCI)
approaches. In Sec.~\ref{sec:Results}, we present a detailed comparison
of our new approaches to established benchmarks, as well as a discussion
of the relative benefits and drawbacks of our approach in comparison
to established methods. Our conclusions are presented in Sec.~\ref{sec:Conclusions}.

\section{Model and methods\label{sec:Model-and-method}}

To avoid redundancy with the companion paper, we only include specific
details needed for the following discussion. In particular, we specify
the particular form of the spectral density and provide a detailed
analysis of the convergence properties for both the DCCI and the SBCI.

\subsection{Spin\textendash boson model and parameters}

We set $\hbar=1$ and consider a spin\textendash boson Hamiltonian
of the form
\begin{equation}
H=\epsilon\hat{\sigma}_{z}+\Delta\hat{\sigma}_{x}+\hat{\sigma}_{z}\sum_{\ell}c_{\ell}x_{\ell}+\sum_{\ell}\frac{1}{2}\left(p_{\ell}^{2}+\omega_{\ell}^{2}x_{\ell}^{2}\right),
\end{equation}
where the $\hat{\sigma}_{i}$ are Pauli matrices describing the spin,
and the parameters $\varepsilon$ and $\Delta$ are called the bias
and diabatic coupling, respectively. The $x_{\ell}$ and $p_{\ell}$
are boson operators. Throughout this work, we specify the system\textendash bath
coupling strength, which determines the $c_{\ell}$ and $\omega_{\ell}$,
by a spectral density that is linear for small $\omega$ and has a
Lorentzian cutoff for large $\omega$:
\begin{equation}
J_{D}\left(\omega\right)=\frac{\lambda}{2}\frac{\omega_{c}\omega}{\omega_{c}^{2}+\omega^{2}}.
\end{equation}
This is known in the literature as the Debye spectral density.\cite{Weiss1999Quantumdissipativesystems}
The reorganization energy, defined as $\lambda=\frac{4}{\pi}\int\frac{J\left(\omega\right)}{\omega}d\omega=2\sum_{\ell}c_{\ell}^{2}/\omega_{\ell}^{2}$,
sets the maximal system\textendash bath coupling strength. The cutoff
frequency of the Lorentzian function, $\omega_{c}$, characterizes
the band width of the bath. Therefore, at equilibrium, the version
of the spin\textendash boson model we consider is fully characterized
by five parameters: the diabatic coupling $\Delta$, the bias energy
$\epsilon$, the cut-off frequency $\omega_{c}$, the reorganization
energy $\lambda$, and the temperature $k_{B}T=1/\beta$ of the boson
bath ($k_{B}$ is Boltzmann constant).

We assume a factorized initial condition given by $\rho_{0}=\rho_{s}\otimes\rho_{b}$.
The initial condition of the spin is $\rho_{s}=\left|1\right\rangle \left\langle 1\right|$,
and the initial density matrix of the bath is thermal equilibrium
in the absence of the system\textendash bath coupling, 
\begin{equation}
\rho_{b}=\frac{e^{-\beta H_{b}}}{\mathrm{Tr}_{b}\left\{ e^{-\beta H_{b}}\right\} }.
\end{equation}
This is where the dependence on (inverse) temperature $\beta$ appears.
We will concentrate on the local dynamics of the spin operator $\sigma_{z}$,
\begin{equation}
\left\langle \sigma_{z}\left(t\right)\right\rangle =\mathrm{Tr}\left\{ \rho_{0}e^{iHt}\hat{\sigma}_{z}e^{-iHt}\right\} ,
\end{equation}
which is often referred to the ``population difference'' of the
spin.

\subsection{Convergence analysis}

To obtain a simple estimate for how rapidly the inchworm approaches
are expected to converge in different regions of parameter space,
we focus on the lowest-order nontrivial contribution in each type
of expansion and determine its magnitude as a function of model parameters.
We consider the $2^{\mathrm{nd}}$-order term of the SBCI and DCCI
expansions, which in both cases can be written in the form
\begin{equation}
G_{2}\left(t\right)=\int_{0}^{t}\mathrm{d}t_{1}\int_{0}^{t_{2}}\mathrm{d}t_{2}C\left(t_{1},t_{2}\right).
\end{equation}
Here, $C\left(t_{1},t_{2}\right)$ is the bath correlation function
associated with each expansion. Loosely speaking, one might expect
an expansion to converge rapidly as long as the corresponding $G_{2}\left(t\right)$
is not significantly greater than unity. Given the functional form
of the Debye spectral density, we can easily evaluate $G_{2}\left(t\right)$
in an analytical or semi-analytical fashion (i.e. by numerical quadrature).

For the SBCI expansion, we can evaluate $G_{2}\left(t\right)$ in
the high and low temperature limits and then derive the convergence
conditions from the appropriate dimensionless parameters that emerge.
This scheme is analogous to one which has been used to determine the
limitations of Redfield theory.\cite{Montoya-Castillo2015ExtendingapplicabilityRedfield}
The bath correlation function in the SBCI expansion is given by $C\left(t_{1},t_{2}\right)=\left\langle \widetilde{B}\left(t_{1}\right)\widetilde{B}\left(t_{2}\right)\right\rangle _{b}$
where $\widetilde{B}\left(t\right)=\sum_{\ell}c_{\ell}\widetilde{x}_{\ell}\left(t\right)$,
and with the definition $\left\langle \cdot\right\rangle _{b}\equiv\mathrm{Tr}_{b}\left\{ \rho_{b}\cdot\right\} $.
The explicit expression is derived in Eq.~(31) of the companion
paper:
\begin{equation}
\begin{split} & \left\langle \widetilde{B}\left(t_{1}\right)\widetilde{B}\left(t_{2}\right)\right\rangle _{b}=\frac{2}{\pi}\int d\omega J_{D}\left(\omega\right)\times\\
 & \ \left[\coth\left(\frac{\beta\omega}{2}\right)\cos\omega\left(t_{1}-t_{2}\right)-i\sin\omega\left(t_{1}-t_{2}\right)\right].
\end{split}
\end{equation}
The integral takes the form $G_{2}\left(t\right)=\xi g\left(t\right)$,
where $\xi$ is a dimensionless parameter and $g\left(t\right)$ is
a time-dependent dimensionless function. We expect the expansion to
converge rapidly as long as $\xi\lesssim1$. In the high temperature
limit $\beta\text{\ensuremath{\omega}}_{c}\ll1$, and we can approximate
$\coth\left(\frac{\beta\omega}{2}\right)\approx\frac{2}{\beta\omega}$
and obtain the dimensionless form for $\xi$: 
\begin{equation}
\xi=\frac{\lambda}{\beta\omega_{c}^{2}}-i\frac{\lambda}{2\omega_{c}}.
\end{equation}
Thus, in this regime, an estimate for the condition for convergence
of the SBCI approach is
\begin{equation}
\frac{\lambda}{\beta\omega_{c}^{2}}\apprle1.\label{eq:high_temp_limit_condition}
\end{equation}
In the low temperature limit $\beta\text{\ensuremath{\omega}}_{c}\gg1$,
we can use $\coth\left(\frac{\beta\omega}{2}\right)\approx1$, but
cannot carry out the integral analytically. In the same spirit, we
factor out the dimensionless scale of the integral, $\mathrm{Re}\left[G_{2}\left(t\right)\right]=\frac{\lambda}{\pi\omega_{c}}\int\mathrm{d}x\frac{1}{x^{2}+1}\frac{1}{x}\left(1-\cos\omega_{c}xt\right)$,
which yields a convergence condition for the low temperature limit
\begin{equation}
\frac{\lambda}{\pi\omega_{c}}\apprle1.\label{eq:low_temp_limit_condition}
\end{equation}
It is noteworthy that since $G_{2}\left(t\right)$ is proportional
to $\lambda\coth\left(\beta\omega/2\right)$, the SBCI expansion becomes
more difficult to converge as $\lambda$ increases or $\beta$ decreases.

The explicit form of the bath correlation function in the DCCI expansion
is given by $C\left(t_{1},t_{2}\right)=e^{-4{\cal Q}_{2}\left(t_{1}-t_{2}\right)-i4{\cal Q}_{1}\left(t_{1}-t_{2}\right)}$,
where 
\begin{equation}
{\cal Q}_{1}\left(t\right)=\frac{2}{\pi}\int\mathrm{d}\omega\frac{J_{D}\left(\omega\right)}{\omega^{2}}\sin\omega t,
\end{equation}
\begin{equation}
{\cal Q}_{2}\left(t\right)=\frac{2}{\pi}\int\mathrm{d}\omega\frac{J_{D}\left(\omega\right)}{\omega^{2}}\coth\left(\frac{\beta\omega}{2}\right)\left(1-\cos\omega t\right).
\end{equation}
Due to the complicated form of these correlation functions, one cannot
obtain an analytical expression to extract a dimensionless scale parameter.
Therefore, we evaluate the integral numerically at a large enough
time for given model parameters. Note that the subsequent discussion
for the convergence condition is obtained by carrying out the integral
numerically. We also note that ${\cal Q}_{1}$ and ${\cal Q}_{2}$
are linearly dependent on $\lambda$, which yields a $1/\lambda^{2}$
dependence for $G_{2}\left(t\right)$. Therefore, the DCCI expansion
becomes easier to converge as $\lambda$ increases. 

\begin{figure*}
\begin{centering}
\begin{minipage}[t]{0.33\textwidth}%
(a) $\beta\Delta=0.5$
\begin{center}
\includegraphics[bb=0bp 0bp 153bp 145bp,clip]{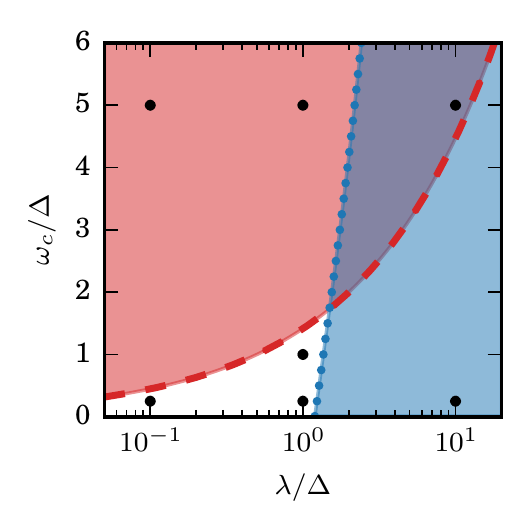}
\par\end{center}%
\end{minipage}%
\begin{minipage}[t]{0.33\textwidth}%
(b) $\beta\Delta=5$
\begin{center}
\includegraphics[bb=0bp 0bp 153bp 145bp,clip]{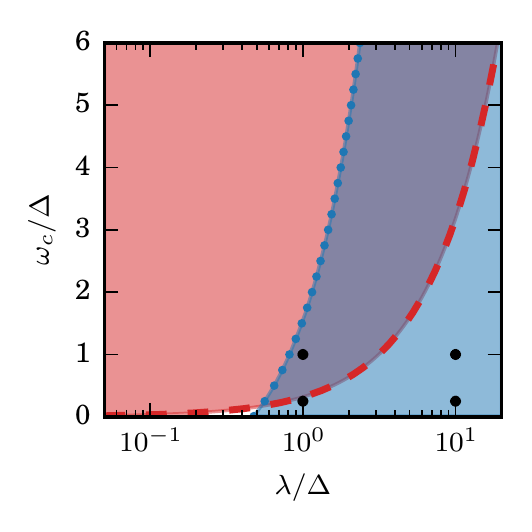}
\par\end{center}%
\end{minipage}%
\begin{minipage}[t]{0.33\textwidth}%
(c) $\beta\Delta=50$
\begin{center}
\includegraphics[bb=0bp 0bp 153bp 145bp,clip]{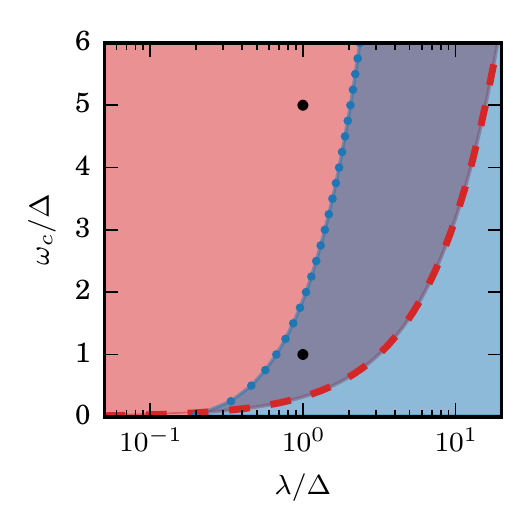}
\par\end{center}%
\end{minipage}
\par\end{centering}
\caption{Spin\textendash boson model parameter space with zero bias $\epsilon=0$.
The $x$-axis is $\lambda/\Delta$ in $\log$ scale and the $y$-axis
is $\omega_{c}/\Delta$ in linear scale. The bath temperatures are
(a) $\beta\Delta=0.5$, (b)$\beta\Delta=5$, and (c) $\beta\Delta=50$.
In each ``phase diagram'', the estimated region of rapid convergence
for the SBCI approach is to the left of the dashed line (red) and
is to the right of the dotted line (blue) for the DCCI approach. Points
indicate the parameters for plots presented in this work.\label{fig:phase_diagram}}
\end{figure*}

A two-dimensional ``phase diagram'' can be drawn as cuts of the
full parameter space with varying $\lambda$ and $\omega_{c}$, shown
in Fig.~\ref{fig:phase_diagram}. Here we limit the discussion to
the subspace with zero energy bias $\epsilon=0$. The horizontal axis
is the scaled reorganization energy ($\lambda/\Delta$) in $\log$
scale and the vertical axis is the scaled cutoff frequency ($\omega_{c}/\Delta$).
Within this coordinate system, we can demarcate the estimated region
of facile convergence for the SBCI and DCCI expansions by the conditions
given above. The red regions indicate the subspace satisfying Eq.~(\ref{eq:high_temp_limit_condition})
and (\ref{eq:low_temp_limit_condition}), in which the SBCI approach
is expected to converge rapidly. The blue regions are obtained by
semi-analytical estimation of the analogous condition for the DCCI
approach.

Fig.~\ref{fig:phase_diagram} exhibits these complementary regions
and shows that their combined area covers much of the relevant parameter
space. We will briefly point out some important features of the phase
diagram. First, for any cutoff frequency $\omega_{c}$, the SBCI
converges better in the small $\lambda$ direction while the DCCI
is expected to work better as $\lambda$ increases. Second, the region
of utility for the SBCI expansion shrinks in the adiabatic regime
(small $\omega_{c}$), which is due to the fact that the correlation
functions in the SBCI expansion have a longer correlation time when
$\omega_{c}$ is small. Lastly, as the temperature decreases, the
regions of rapid convergence of both the SBCI and DCCI approaches
expand and cover almost the entire parameter space.

While Fig.~\ref{fig:phase_diagram} provides an illustration of
applicable regions of our approach, the regions are determined by
rough estimation of lowest order contribution. In principle, our inchworm
expansions are numerically exact in the entire parameter space, as
discussed in the companion paper. In the ``uncovered'' or white
region, our approaches should continue to yield reliable dynamical
behavior at least on some time scales, albeit with potentially much
greater numerical effort.

\section{Results\label{sec:Results}}

\subsection{Computational Methodology}

\begin{figure}
\begin{raggedright}
(a) $\lambda/\Delta=10$, $\omega_{c}/\Delta=5$
\par\end{raggedright}
\begin{centering}
\includegraphics{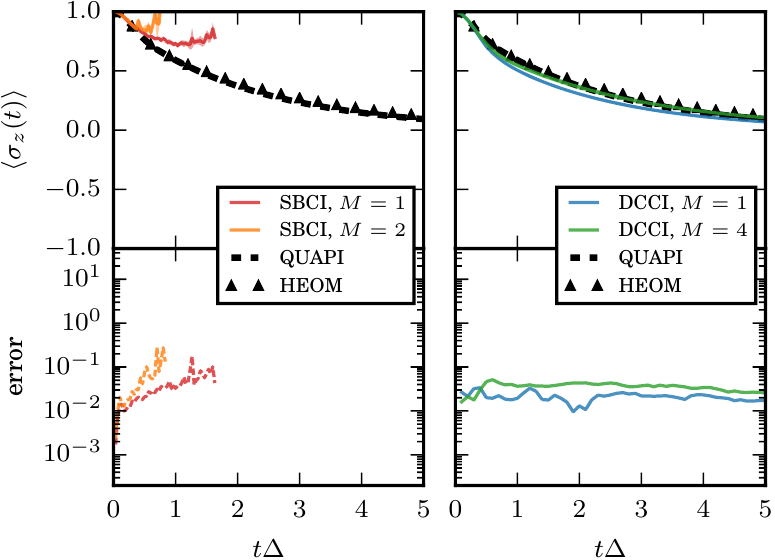}
\par\end{centering}
\begin{raggedright}
(b) $\lambda/\Delta=10$, $\omega_{c}/\Delta=0.25$
\par\end{raggedright}
\begin{centering}
\includegraphics{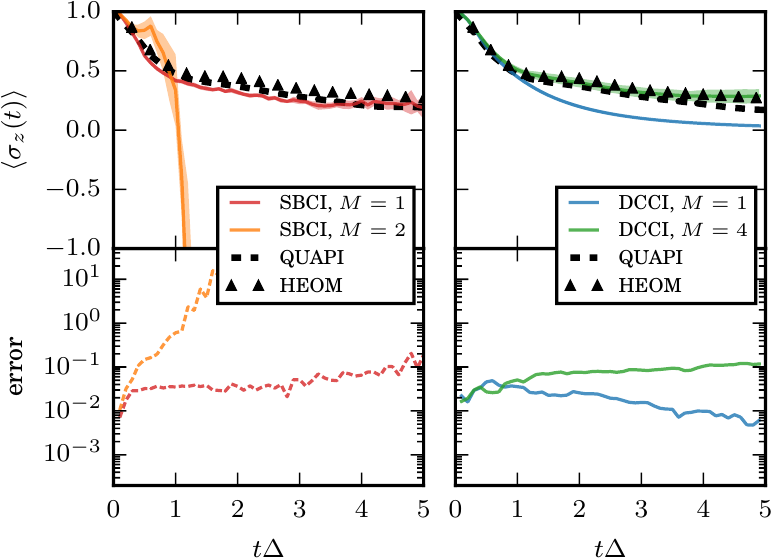}
\par\end{centering}
\caption{Nonequilibrium population difference $\left\langle \sigma_{z}\left(t\right)\right\rangle $
(top subplots) and corresponding error estimates (bottom subplots)
as a function of time in the strong coupling ($\lambda/\Delta=10$)
and high temperature ($\beta\Delta=0.5$) regime. The bias energy
is $\epsilon=0$. The results calculated by the SBCI (left panels,
red and orange) and DCCI (right panels, blue and green) inchworm expansions
are plotted for (a) a non-adiabatic (fast) bath with $\omega_{c}/\Delta=5$,
and (b) an adiabatic (slow) bath with $\omega_{c}/\Delta=0.25$. The
time step of SBCI for (a) is $\mathrm{d}t\Delta=0.1/3$. The error
estimate of the SBCI calculation (dashed lines in red and organe)
is for one single run. Maximum order for a inchworm step is indicated
by $M$. The thickness of the Monte Carlo results results from our
error estimates. The dashed lines are the QUAPI results with (a)$\Delta t=0.1$,
$k_{\mathrm{max}}=6$ and (b)$\Delta t=0.3$, $k_{\mathrm{max}}=11$.
The triangles indicate the HEOM result with $K=2$ and $L=20$. \label{fig:epsilon0.0_delta1.0_lambda10_omega_c*_beta0.5}}
\end{figure}

The general framework of the two general types of inchworm expansions
used here may be found in the companion paper. In the following,
each inchworm step is limited to a fixed run time and the order of
each individual inchworm diagram is restricted to a maximum order
$M$. We use $\mathrm{d}t\Delta=0.1$ for the size of the inchworm
step in the following calculation, unless otherwise specified. One
may then check for convergence by systematically increasing $M$,
decreasing $\mathrm{d}t$ and increasing the number of Monte Carlo
samples.\cite{Cohen2015Tamingdynamicalsign} The SBCI calculation
requires the full information contained in two-time restricted propagators,
thus for the SBCI propagation to a time $t=N\mathrm{d}t$ requires
$N^{2}$ inchworm steps (in fact, by taking advantage of time-reversal
symmetry and the contour ordering of the time arguments, the number
of steps needed turns out to be $\sim\frac{1}{4}N^{2}$). On the other
hand, the DCCI expansion is phrased solely in terms of single-time
properties, such that it requires only $N$ inchworm steps to reach
a simulation time $t=N\mathrm{d}t$. For both approaches, we perform
multiple independent inchworm calculations in order to properly account
for error propagation.\cite{Cohen2015Tamingdynamicalsign} Note that
the error propagation shown in the lower subplot of all the figures
is this ``full'' error, except for the left panels in Fig~\ref{fig:epsilon0.0_delta1.0_lambda10_omega_c*_beta0.5},
where the standard Monte Carlo statistical error is shown for didactic
purposes.

We compare our calculations with several existing numerically exact
methods, including the quasi-adiabatic propagator path integral (QUAPI),\cite{Makarov1994Pathintegralsdissipative,Makri1995Numericalpathintegral,Makri1995Tensorpropagatoriterative,Makri1996Longtimequantum}
hierarchical equations of motion (HEOM) method,\cite{Tanimura1989TimeEvolutionQuantum,Ishizaki2009Unifiedtreatmentquantum,Struempfer2012OpenQuantumDynamics}
and the multi-configuration time-dependent Hartree (MCTDH) approach.\cite{Thoss2001Selfconsistenthybrid,Wang2001Systematicconvergencein,Wang2003Multilayerformulationmulticonfiguration}
QUAPI is based on the discretization of influence functional for
reduced propagation on the Keldysh contour. The maximum number of
short-time propagators that the path integral spans is determined
by a parameter $k_{\mathrm{max}}$, which governs the memory length.
The approach becomes difficult to converge when the memory length
is long. The HEOM approach introduces a hierarchy of auxiliary density
matrices and employs a Mastsubara expansion for the bath density matrix.
The hierarchy truncation level $L$ and number of Matsubara terms
$K$ are numerical parameters that are tuned to converge the HEOM
calculation. A standard, highly parallel implementation is available,\cite{Struempfer2012OpenQuantumDynamics}
known to be accurate in the high temperature limit and for the Debye
spectral density. Generically, the HEOM approach has more difficulty
for low temperatures and non-Debye spectral densities. The MCTDH
approach is based on the expansion of the interacting many-body wave
function as a tensor product of wavefunctions defined in a convenient
set of orbitals. A highly efficient protocol may then be used to control,
in a time-dependent manner, the number of orbitals needed for exact
convergence. Exact MCTDH results for the spin\textendash boson model
are reported in Ref.~\onlinecite{Thoss2001Selfconsistenthybrid}.

We will be using the benchmarks to investigate accuracy, and will
make no attempt to compare numerical efficiency beyond general points
having to with the computational scaling of the algorithms. To provide
a general idea, we will say that using our current implementation,
most of the (linear scaling in time) DCCI results presented here can
be comfortably obtained on a laptop in minutes to hours, whereas the
(quadratically scaling in time) SBCI results typically require a small
cluster. However, it should be noted that the data below was obtained
with a very flexible but not at all optimized code written in the
high-level Python programming language. From our experience with similar
algorithms for the Anderson impurity model,\cite{Cohen2015Tamingdynamicalsign}
we estimate that 1\textendash 2 orders of magnitude in overall runtime
could be achieved by writing an efficient code, or simply by switching
to a compiled language.

\subsection{High temperature regime\label{subsec:High-Temperature-and}}

\begin{figure}
\begin{raggedright}
(a) $\lambda/\Delta=0.1$, $\omega_{c}/\Delta=5$
\par\end{raggedright}
\begin{centering}
\includegraphics{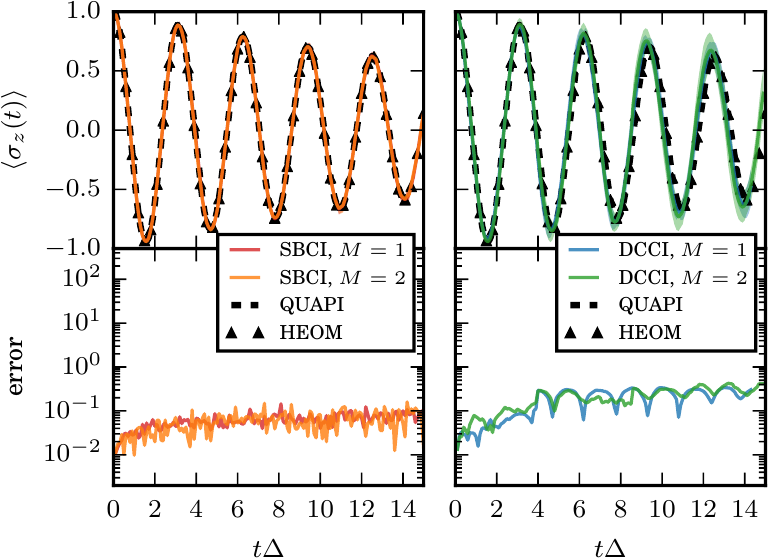}
\par\end{centering}
\begin{raggedright}
(b) $\lambda/\Delta=0.1$, $\omega_{c}/\Delta=0.25$
\par\end{raggedright}
\begin{centering}
\includegraphics{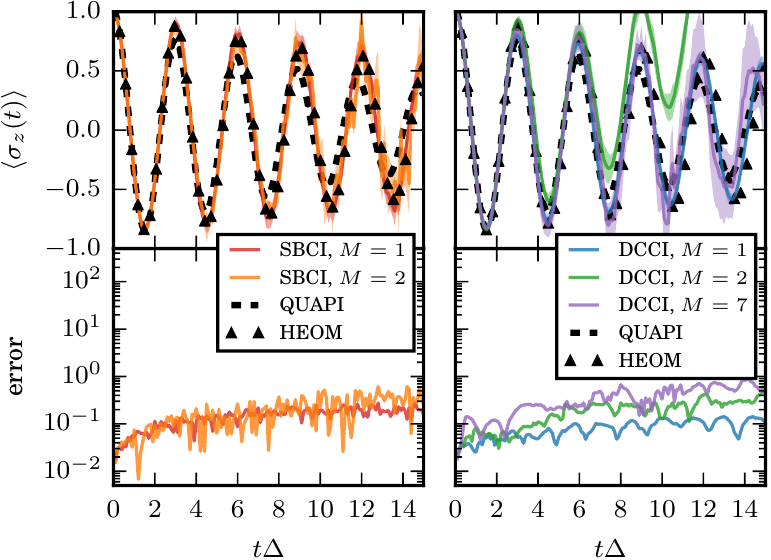}
\par\end{centering}
\caption{Nonequilibrium population difference $\left\langle \sigma_{z}\left(t\right)\right\rangle $
(top subplots) and corresponding error estimates (bottom subplots)
as a function of time in the weak coupling ($\lambda/\Delta=0.1$)
and high temperature ($\beta\Delta=0.5$) regime. The bias energy
is $\epsilon=0$. The results calculated by the SBCI (left panels,
red and orange) and DCCI (right panels, blue and green) inchworm expansions
are plotted for (a) a non-adiabatic (fast) bath with $\omega_{c}/\Delta=5$,
and (b) an adiabatic (slow) bath with $\omega_{c}/\Delta=0.25$. Maximum
order for an inchworm step is indicated by $M$. The thickness of
the Monte Carlo results results from our error estimates. The dashed
lines are the QUAPI results with (a)$\Delta t=0.1$, $k_{\mathrm{max}}=6$
and (b)$\Delta t=0.1$, $k_{\mathrm{max}}=12$. The triangles indicate
the HEOM result with $K=2$ and $L=20$. \label{fig:epsilon0.0_delta1.0_lambda0.1_omega_c*_beta0.5}}
\end{figure}

\begin{figure}
\begin{raggedright}
(a) $\lambda/\Delta=1$, $\omega_{c}/\Delta=5$
\par\end{raggedright}
\begin{centering}
\includegraphics{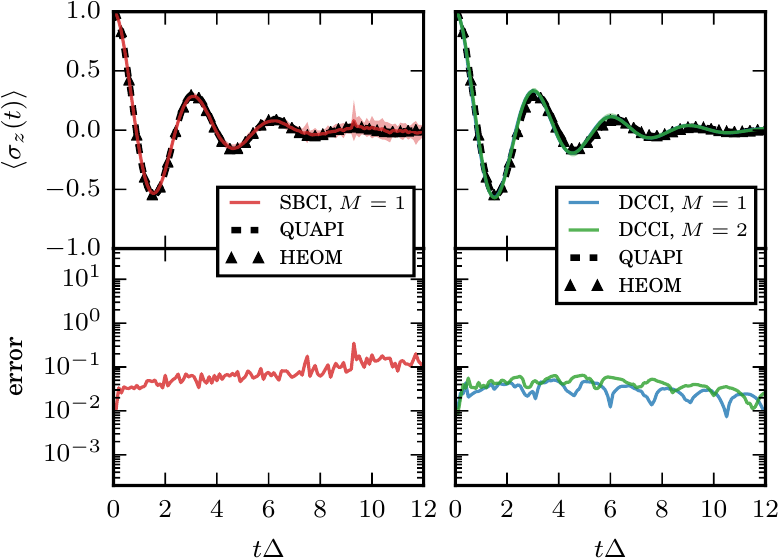}
\par\end{centering}
\begin{raggedright}
(b) $\lambda/\Delta=1$, $\omega_{c}/\Delta=1$
\par\end{raggedright}
\begin{centering}
\includegraphics{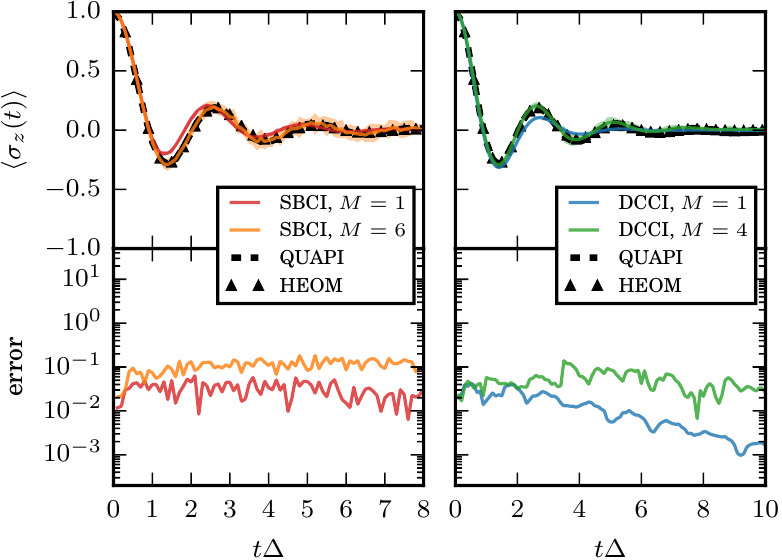}
\par\end{centering}
\begin{raggedright}
(c) $\lambda/\Delta=1$, $\omega_{c}/\Delta=0.25$
\par\end{raggedright}
\begin{centering}
\includegraphics{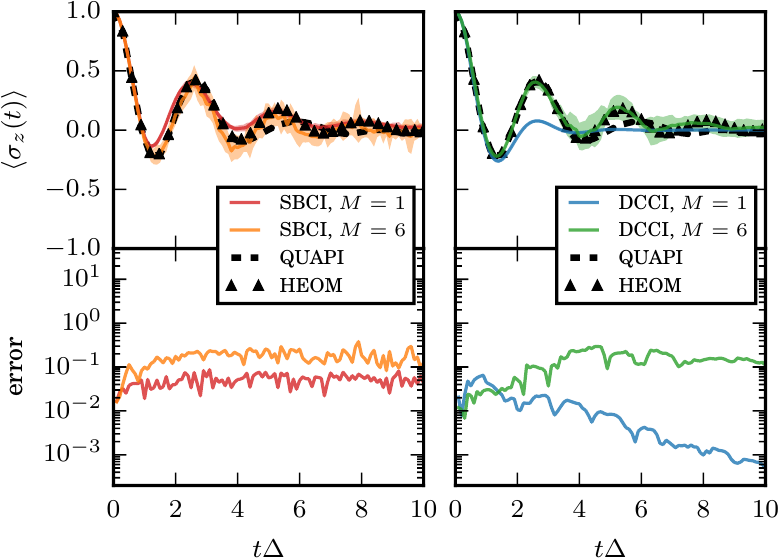}
\par\end{centering}
\caption{Nonequilibrium Population difference $\left\langle \sigma_{z}\left(t\right)\right\rangle $
(top subplots) and corresponding error estimates (bottom subplots)
as a function of time in the intermediate coupling ($\lambda/\Delta=1$)
and high temperature ($\beta\Delta=0.5$) regime. The bias energy
is $\epsilon=0$. The results calculated by the SBCI (left panels,
red and orange) and DCCI (right panels, blue and green) expansions
are plotted for (a) a non-adiabatic (fast) bath with $\omega_{c}/\Delta=5$,
(b) an intermediate bath with $\omega_{c}/\Delta=1$, and (c) an adiabatic
(slow) bath with $\omega_{c}/\Delta=0.25$. Maximum order for a inchworm
step is indicated by $M$. The thickness of the Monte Carlo results
results from our error estimates. The dashed line are the QUAPI results
with (a) $\Delta t=0.1$, $k_{\mathrm{max}}=6$, (b) $\Delta t=0.2$,
$k_{\mathrm{max}}=10$, and (c) $\Delta t=0.3$, $k_{\mathrm{max}}=11$.
The triangles indicate the HEOM result with $K=2$ and $L=20$. \label{fig:epsilon0.0_delta1.0_lambda1_omega_c*_beta0.5}}
\end{figure}
We start our comparison of the inchworm approaches with other exact
methods in the high temperature regime (Fig.~(\ref{fig:phase_diagram})(a)),
specifically $\beta\Delta=0.5$ ($k_{B}T/\Delta=2$), and consider
the vertical cuts at weak coupling $\lambda/\Delta=0.1$, strong coupling
$\lambda/\Delta=10$, and intermediate coupling $\lambda/\Delta=1$
in the following.

\subsubsection{Weak coupling}

In the weak system\textendash bath coupling regime, we consider cases
with scaled reorganization energy ($\lambda/\Delta=0.1$) where we
expect the SBCI expansion to converge more easily than the DCCI expansion.
In Fig.~\ref{fig:epsilon0.0_delta1.0_lambda0.1_omega_c*_beta0.5},
we find that the lowest order ($M=1$) results for the SBCI expansion
always gives a quantitative account of the dynamics with the error
remaining nearly constant over the simulation time. The SBCI result
also converges rapidly upon increasing the maximum order $M$ of each
inchworm step. We note that a smaller cut-off frequency yields a greater
statistical error (see the lower panels of Fig.~\ref{fig:epsilon0.0_delta1.0_lambda0.1_omega_c*_beta0.5}(a)
and (b)) with the same computational cost. This is due to the long
correlation time induced by a small $\omega_{c}$, which makes it
more difficult to converge the SBCI expansion. On the other hand,
the DCCI calculation also yields surprisingly accurate results. However,
for a small cut-off frequency, it becomes more difficult to converge
the DCCI approach, as can be seen in the right panel of Fig.\ref{fig:epsilon0.0_delta1.0_lambda0.1_omega_c*_beta0.5}(b).
Note that for the DCCI approach, convergence is non-monotonic: whereas
the $M=1$ case appears quite accurate, the $M=2$ case actually yields
results that are substantially worse. To overcome this, it is necessary
to include much higher order inchworm diagrams. In this case $M=7$
appears to be sufficient. When many contributions at high orders are
important, a larger investment of computer time is generally required
to overcome the sign problem (here, \textasciitilde{}3 times more
computing resources were used to obtain the still noisy $M=7$ result
than was used for $M=1$ and $M=2$).

Thus, in the high temperature, weak coupling regime, both inchworm
approaches appear capable of reproducing the results obtained by the
HEOM method, which easily converges to the exact answer for the Debye
spectral density at high temperatures. The DCCI approach does show
some convergence difficulties in this regime for the slow bath case.
We could not converge QUAPI in the slow bath regime, and quantitative
discrepancies can be found between QUAPI and HEOM here, as seen in
Fig.~\ref{fig:epsilon0.0_delta1.0_lambda0.1_omega_c*_beta0.5}.

\subsubsection{Strong coupling}

For strong system\textendash bath coupling ($\lambda/\Delta=10$),
we anticipate that the SBCI expansion will be difficult to converge
and the DCCI expansion will show rapid convergence. The right panels
of Fig.~\ref{fig:epsilon0.0_delta1.0_lambda10_omega_c*_beta0.5}(a)
and (b) show that the DCCI results converge to accurate population
dynamics as we increase the maximum order $M$ of each inchworm step,
but that at least $M=4$ is required for convergence. As expected,
the SBCI expansion is difficult to converge in this parameter regime.
The origins of this convergence issue can be gleaned from the behavior
of the error estimate. In particular, the error estimates found in
the left panels in Fig.~\ref{fig:epsilon0.0_delta1.0_lambda10_omega_c*_beta0.5}
show the statistical error for \emph{one single} SBCI calculation,
which indicates the error of the Monte Carlo estimation of the integral
within each inchworm step. This is an underestimate of the error margin,
as it does not take into account the error propagation from shorter
times; other plots in this paper show the full error analysis. We
note that even the single run error increases exponentially with time,
so that it is clear that the origin of the exponential growth in noise
to signal ratio is actually the sign problem and not error propagation.
The weight of high order configurations to the integral becomes large
when $\lambda$ increases, and the SBCI expansion could not be converged
in Fig.~\ref{fig:epsilon0.0_delta1.0_lambda10_omega_c*_beta0.5}(a)
even with a third of the inchworm step size used in the rest of the
figures in this paper. To capture these high order contributions,
one may increase $M$. However, as shown in Fig.~\ref{fig:epsilon0.0_delta1.0_lambda10_omega_c*_beta0.5},
the slope of the statistical error grows unfavorably in this case
as we increase $M$, rendering the SBCI expansion difficult to converge.
This is the only part of the parameter space where we have found that
one of the two methods proposed in the companion paper completely
fails to overcome the dynamical sign problem.

We also note that the full error may not always monotonically increase
with time and can exhibit decreasing or non-monotonic behavior. This
is due to the fact that all contributions are dressed with \emph{full
propagators} tailored to the observable being measured. Where the
value of the observable is small, each and every contribution will
therefore be composed of small propagators and tend to be accordingly
small, such that the statistical noise obtained is proportional to
the (absolute) value at the point being measured.

\subsubsection{Intermediate coupling}

Lastly, we focus on the intermediate system\textendash bath coupling
regime where the scaled reorganization energy is $\lambda/\Delta=1$.
Fig.~\ref{fig:epsilon0.0_delta1.0_lambda1_omega_c*_beta0.5} exhibits
a general feature of the inchworm approaches: convergence with respect
to the maximum order becomes more difficult to obtain as the cut-off
frequency decreases. For a fast bath ($\omega_{c}/\Delta=5$), both
the SBCI and DCCI expansions yield quite accurate results at lowest
order. For $\omega_{c}/\Delta=1$, the parameter set $\left(\lambda/\Delta,\omega_{c}/\Delta\right)=\left(1,1\right)$
is located outside of the ``safe'' regions for the SBCI and DCCI
as demarcated in Fig.~\ref{fig:phase_diagram}(a). Here we observe
clear, but small, discrepancies between the SBCI/DCCI results for
$M=1$ and numerically exact dynamics. By systematically increasing
$M$, the discrepancies can be corrected. When the cut-off frequency
is small, the parameter set $\left(\lambda/\Delta,\omega_{c}/\Delta\right)=\left(1,0.25\right)$
is particularly difficult for both SBCI and DCCI expansions, although
convergence is still seen for $M=6$. Lastly, note that here, as in
Fig.~\ref{fig:epsilon0.0_delta1.0_lambda10_omega_c*_beta0.5}, some
notable discrepancies exist between the HEOM and QUAPI results. The
inchworm expansions always converge to the HEOM results, which are
expected to be more reliable in the high temperature regime.

\subsection{Low temperature regime\label{subsec:Low-Temperature-and}}

We now turn our attention to the phase diagram in the low temperature
regime, specifically $\beta\Delta=5$ ($k_{B}T/\Delta=0.2$), and
concentrate on vertical cuts at intermediate coupling $\lambda/\Delta=1$
and strong coupling $\lambda/\Delta=10$, using the more suitable
of the two methods in each case. These parameters correspond to Fig.~\ref{fig:phase_diagram}(b).

\subsubsection{Intermediate coupling}

\begin{figure}
\begin{raggedright}
\par\end{raggedright}
\begin{raggedright}
\par\end{raggedright}
\begin{raggedright}
\hspace{1cm}(a)$\lambda/\Delta=1$, $\omega_{c}/\Delta=1$\hspace{0.5cm}(b)$\lambda/\Delta=1$,
$\omega_{c}/\Delta=0.25$
\par\end{raggedright}
\begin{centering}
\includegraphics{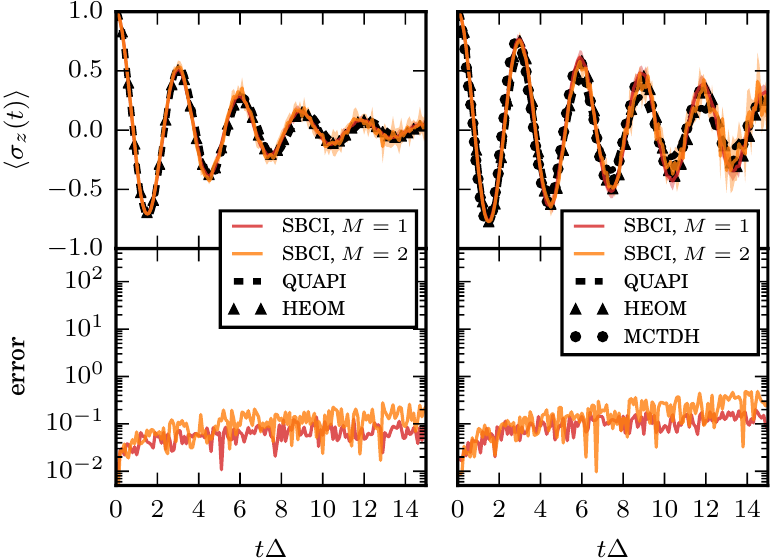}
\par\end{centering}
\caption{Nonequilibrium Population difference $\left\langle \sigma_{z}\left(t\right)\right\rangle $
(top subplots) and corresponding error estimates (bottom subplots)
as a function of time in the intermediate coupling ($\lambda/\Delta=1$)
and low temperature ($\beta\Delta=5$) regime. The bias energy is
$\epsilon=0$. The results calculated by the SBCI (red lines) expansions
are plotted for (a) an intermediate bath with $\omega_{c}/\Delta=1$
and (b) an adiabatic bath with $\omega_{c}/\Delta=0.25$. The maximum
order for the inchworm step shown is $M=1$. The thickness of the
Monte Carlo results results from our error estimates. The dashed lines
are the QUAPI results with (a) $\Delta t=0.1$, $k_{\mathrm{max}}=6$
and (b) $\Delta t=0.1$, $k_{\mathrm{max}}=10$. The triangles indicate
the HEOM result with $K=2$ and $L=20$. The MCTDH data is reported
in Ref.~\onlinecite{Thoss2001Selfconsistenthybrid}. \label{fig:epsilon0.0_delta1.0_lambda1_omega_c*_beta5.0}}
\end{figure}

For intermediate coupling strength ($\lambda/\Delta=1$), the SBCI
expansion is expected to converge at low temperatures more easily
than in the high temperature regime. In particular, Fig.~\ref{fig:phase_diagram}
shows the region of rapid convergence for the SBCI expansion becomes
larger at low temperatures (b) than high temperatures (a). In Fig.~\ref{fig:epsilon0.0_delta1.0_lambda1_omega_c*_beta5.0},
we find that the SBCI expansion can provide accurate results even
at $M=1$ for the parameter sets $\left(\lambda/\Delta,\omega_{c}/\Delta\right)=\left(1,1\right)$
and $\left(1,0.25\right)$, which would be more difficult to converge
in the high temperature regime discussed in Sec.~\ref{subsec:High-Temperature-and}. 

\subsubsection{Strong coupling}

\begin{figure}
\begin{raggedright}
\par\end{raggedright}
\begin{raggedright}
\hspace{0.9cm}(a)$\lambda/\Delta=10$, $\omega_{c}/\Delta=1$\hspace{0.6cm}(b)$\lambda/\Delta=10$,
$\omega_{c}/\Delta=0.25$
\par\end{raggedright}
\begin{centering}
\includegraphics{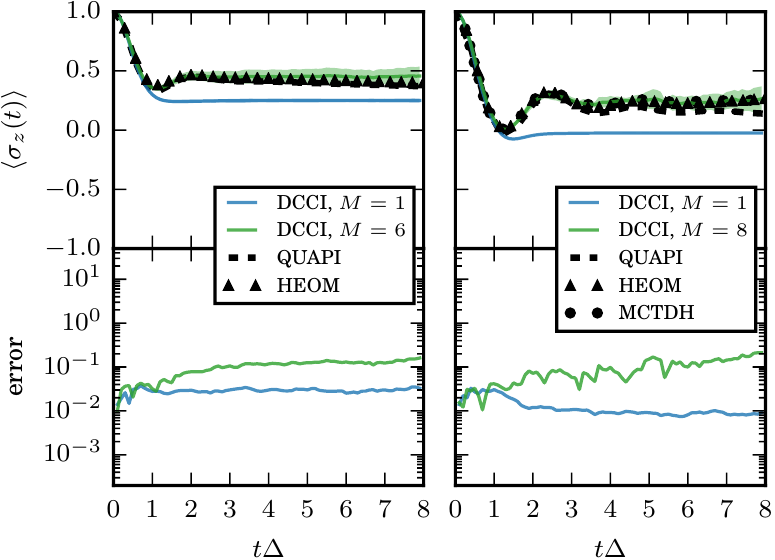}
\par\end{centering}
\caption{Nonequilibrium Population difference $\left\langle \sigma_{z}\left(t\right)\right\rangle $
(top subplots) and corresponding error estimates (bottom subplots)
as a function of time in the strong coupling ($\lambda/\Delta=10$)
and low temperature ($\beta\Delta=5$) regime. The bias energy is
$\epsilon=0$. The results calculated by the DCCI (blue and green
lines) expansions are plotted for (a) an intermediate bath with $\omega_{c}/\Delta=1$
and (b) an adiabatic bath with $\omega_{c}/\Delta=0.25$. Maximum
order for a inchworm step is indicated by $M$. The thickness of the
Monte Carlo results results from our error estimates. The dashed line
are the QUAPI results with (a) $\Delta t=0.2$, $k_{\mathrm{max}}=11$
and (b) $\Delta t=0.4$, $k_{\mathrm{max}}=10$. The triangles indicate
the HEOM result with $K=3$ and $L=20$. The MCTDH data is reported
in Ref.~\onlinecite{Thoss2001Selfconsistenthybrid}.\label{fig:epsilon0.0_delta1.0_lambda10_omega_c*_beta5.0}}
\end{figure}

In the strong coupling regime ($\lambda/\Delta=10$), the DCCI approach
is more rapidly convergent and efficient than the SBCI expansion (see
Fig.~\ref{fig:epsilon0.0_delta1.0_lambda10_omega_c*_beta5.0}). In
particular, we show the DCCI results for parameter sets in the adiabatic
and intermediate regime, namely $\left(\lambda/\Delta,\omega_{c}/\Delta\right)=\left(10,1\right)$
and $\left(10,0.25\right)$. In these regimes, the lowest order DCCI
results tend to over-estimate the incoherent decay of the population.
Including higher order contributions within each inchworm step is
necessary, as it provides significant corrections to population dynamics
leading to agreement with the HEOM and MCTDH results. At small $\omega_{c}/\Delta$,
one needs to go as far as $M=8$, which is too difficult to fully
converge with our current prototype code without spending a great
deal of computer time. In the adiabatic regime (small $\omega_{c}$),
QUAPI also tends to overestimate the decay for the long time behavior.
To obtain correct long-time dynamics, one would need to increase the
truncation of the memory length $k_{\mathrm{max}}$, which greatly
increases the need for memory and makes QUAPI difficult to converge.

\subsection{Very low temperature limit}

Finally, we explore the very low temperature limit $\beta\Delta=50$
($k_{B}T/\Delta=0.02$) corresponding to the phase diagram Fig.~\ref{fig:phase_diagram}(c).
In this limit, the standard HEOM implementation\cite{Struempfer2012OpenQuantumDynamics}
can be computationally expensive to converge. Indeed, the lower the
bath temperature, the more Matsubara terms that are needed to capture
the bath density matrix and the more hierarchical levels are required
to converge the long-time dynamics. We find that the HEOM implementation
available to us becomes unfeasible for very low temperatures, though
we note that recent advances may ameliorate this problem in at least
some instances.\cite{Hu2010CommunicationPadespectrum,Hu2011Padespectrumdecompositions,Yan2014Theoryopenquantum,Moix2013hybridstochastichierarchy,Tang2015Extendedhierarchyequation,Ye2016HEOMQUICKprogram}
At least in the Anderson model, this is not always the case.\cite{Haertle2015TransportthroughAnderson}
Fig.~\ref{fig:phase_diagram} suggests that the SBCI and DCCI expansions
hold an advantage over HEOM (though not MCTDH) at low temperatures,
in that the computational cost does not increase with decreasing temperature.
However, since at low enough temperatures strong correlation effects
may alter the picture, it is not trivial that the simple analysis
used to generate this figure should hold.

\begin{figure}
\begin{raggedright}
(a) $\lambda/\Delta=1$, $\omega_{c}/\Delta=5$, $\beta\Delta=50$
\par\end{raggedright}
\begin{centering}
\includegraphics{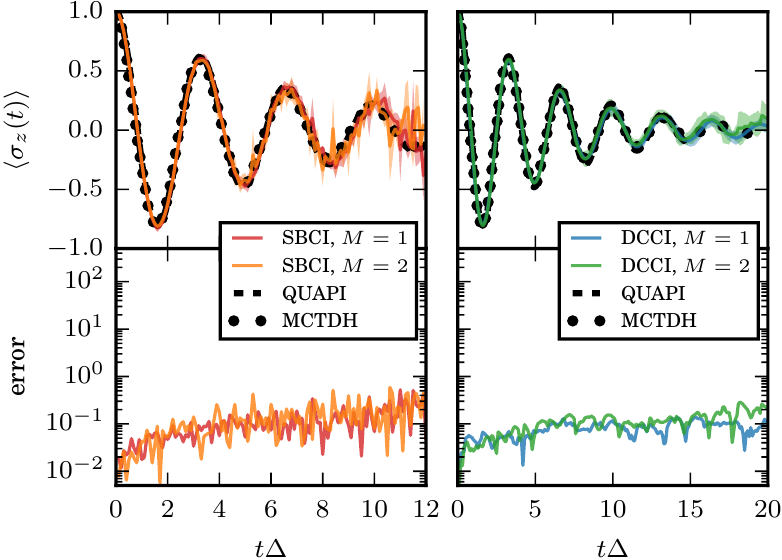}
\par\end{centering}
\begin{raggedright}
(b) $\lambda/\Delta=1$, $\omega_{c}/\Delta=1$, $\beta\Delta=50$
\par\end{raggedright}
\begin{centering}
\includegraphics{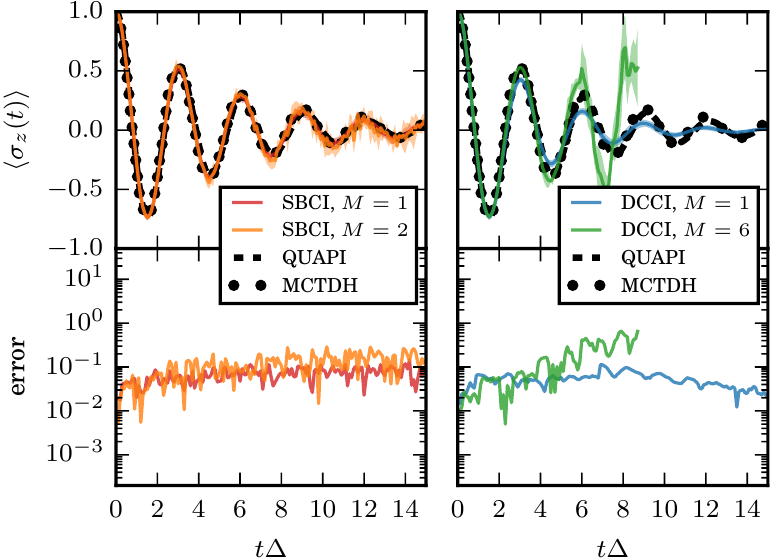}
\par\end{centering}
\caption{Nonequilibrium population difference $\left\langle \sigma_{z}\left(t\right)\right\rangle $
(top subplots) and corresponding error estimates (bottom subplots)
as a function of time in the intermediate coupling ($\lambda/\Delta=1$)
and very low temperature ($\beta\Delta=50$) regime. The bias energy
is $\epsilon=0$. The results calculated by the SBCI (left panels)
and the DCCI (right panels) expansions are plotted for (a) a non-adiabatic
(fast) bath with $\omega_{c}/\Delta=5$ and (b) an intermediate bath
with $\omega_{c}/\Delta=1$. Maximum order for each inchworm step
is indicated by $M$. The thickness of the Monte Carlo results results
from our error estimates. The dashed line are the QUAPI results with
$\Delta t=0.1$ and $k_{\mathrm{max}}=10$. The MCTDH data is reported
in Ref.~\onlinecite{Thoss2001Selfconsistenthybrid}.\label{fig:epsilon0.0_delta1.0_lambda1.0_omega_c*_beta50.0}}
\end{figure}

In Fig.~\ref{fig:phase_diagram}(c), the combined area of strong
convergence for the SBCI and DCCI expansions covers almost the entire
parameter space in the very low temperature case. For the fast bath
case ($\omega_{c}/\Delta=5$), Fig.~\ref{fig:epsilon0.0_delta1.0_lambda1.0_omega_c*_beta50.0}(a)
shows that the parameter set falls out of the region of facile convergence
for the DCCI approach, however we find that the DCCI expansion can
still provide accurate population dynamics, and is in fact more efficient
than the SBCI expansion. On the other hand, for the intermediate cut-off
frequency case ($\omega_{c}/\Delta=1$), the SBCI expansion results
in population dynamics that agree perfectly with the MCTDH result,
while we note that the DCCI expansion is difficult to converge with
respect to the maximum order $M$ (see Fig.~\ref{fig:epsilon0.0_delta1.0_lambda1.0_omega_c*_beta50.0}(b)).
This clearly does not agree with our naive estimates for convergence
of the DCCI expansion.

\subsection{Biased systems\label{subsec:Biased-system}}

\begin{figure}
\begin{raggedright}
(a) $\lambda/\Delta=1$, $\omega_{c}/\Delta=5$, $\beta\Delta=0.5$
\par\end{raggedright}
\begin{centering}
\includegraphics{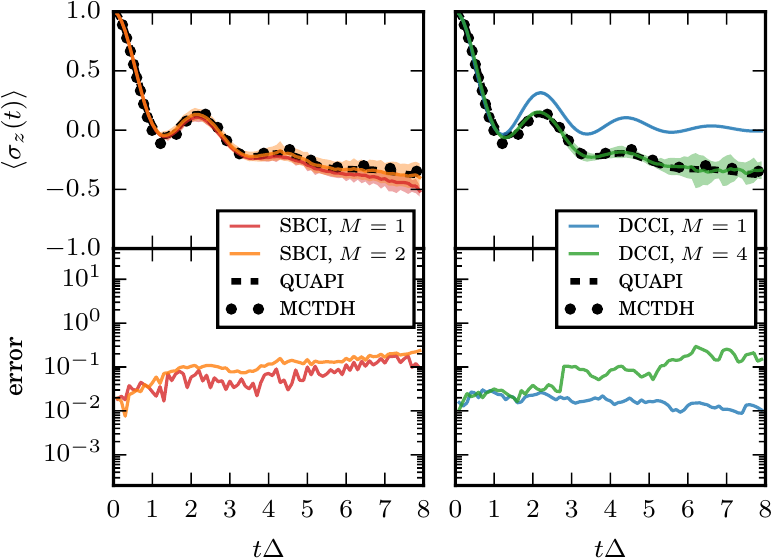}
\par\end{centering}
\begin{raggedright}
(b) $\lambda/\Delta=1$, $\omega_{c}/\Delta=5$, $\beta\Delta=50$
\par\end{raggedright}
\begin{centering}
\includegraphics{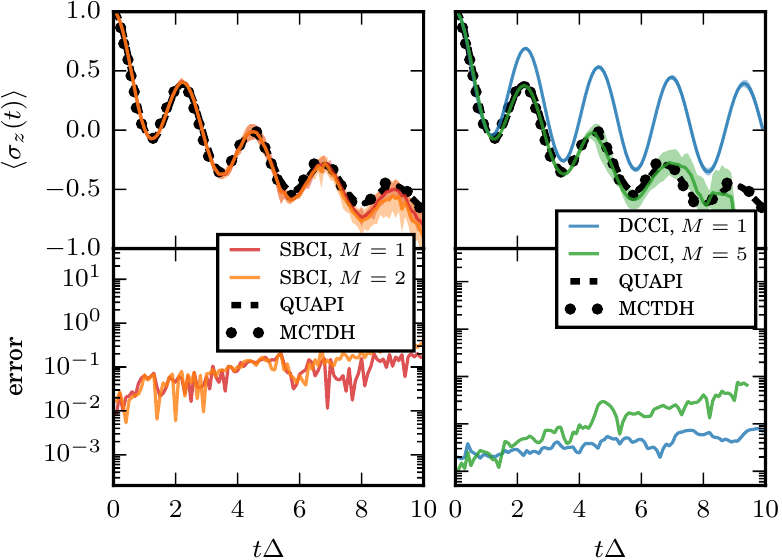}
\par\end{centering}
\caption{Nonequilibrium population difference $\left\langle \sigma_{z}\left(t\right)\right\rangle $
(top subplots) and corresponding error estimates (bottom subplots)
as a function of time in the intermediate coupling ($\lambda/\Delta=1$)
and non-adiabatic ($\omega_{c}/\Delta=5$) regime. The bias energy
is $\epsilon=\Delta$. The results calculated by the SBCI (left panels)
and the DCCI (right panels) expansions are plotted for (a) high temperature
with $\beta\Delta=0.5$ and (b) very low temperature with $\beta\Delta=50$.
Maximum order for each inchworm step is indicated by $M$. The thickness
of the Monte Carlo results results from our error estimates. The dashed
line are the QUAPI results with $\Delta t=0.1$ and $k_{\mathrm{max}}=10$.
The MCTDH data is reported in Ref.~\onlinecite{Thoss2001Selfconsistenthybrid}.\label{fig:epsilon1.0_delta1.0_lambda1.0_omega_c5.0_beta*}}
\end{figure}

We now turn to a discussion of the last dimension of the parameter
space of the spin\textendash boson model, namely the bias energy $\epsilon$
of the spin subsystem. We expect the SBCI and DCCI approaches to have
similar behavior with respect to convergence within parameter space
for non-zero bias energy. However, non-zero bias energy may introduce
an additional phase in the reduced propagators and cause a more difficult
dynamical sign problem. 

For the SBCI expansion, the $\epsilon$ dependence is only found
within the system influence functional: in particular, in the interaction
picture operators $\widetilde{\sigma}_{z}\left(t\right)=e^{iH_{s}t}\hat{\sigma}_{z}e^{-iH_{s}t}$,
where $H_{s}=\epsilon\hat{\sigma}_{z}+\Delta\hat{\sigma}_{x}$. The
bath influence functional does not depend on the bias energy, so that
the inchworm propriety of any individual diagram remains unchanged.
Therefore, it is straightforward to account for the bias energy within
the SBCI algorithm. On the other hand, the DCCI algorithm only contains
$\epsilon$-dependence in the phase influence functional 
\begin{equation}
\Phi\left(\boldsymbol{s}\right)\propto\exp\left[-i\sum_{k=1}^{m+1}\epsilon\sigma_{k}\left(s_{k}-s_{k-1}\right)\right]
\end{equation}
where we designate the state between $\left[s_{k-1,}s_{k}\right]$
as $\sigma_{k}$ for $k\in\left\{ 1,\ldots,m\right\} $ as in the
companion paper. Note that the phase functional is a real number only
if $\epsilon=0$, while $\epsilon\neq0$ renders $\Phi\left(s\right)$
complex and thus potentially increases the dynamical sign problem
of the dQMC method, making the DCCI algorithm somewhat more difficult
to converge.

In Fig.~\ref{fig:epsilon1.0_delta1.0_lambda1.0_omega_c5.0_beta*},
we show the SBCI and DCCI results for non-zero bias energy $\epsilon=\Delta$
cases at high temperature $\beta\Delta=0.5$ and the very low temperature
case $\beta\Delta=50$. The system\textendash bath coupling is taken
to be $\lambda/\Delta=1$ and a cut-off frequency of $\omega_{c}/\Delta=5$
is used. In general, the error estimate of the SBCI expansion grows
more rapidly with time, so that more computational effort to control
the error propagation is needed. The DCCI expansion shows a clear
convergence with respect to the maximum order $M$. Compared to the
same parameter set $\lambda/\Delta=1$, $\omega_{c}/\Delta=5$, and
$\beta\Delta=50$ for zero bias energy, we note that the non-zero
bias energy does increase the computational effort, especially for
the DCCI approach.

\section{Conclusions\label{sec:Conclusions}}

In this work we have presented benchmark calculations of the inchworm
dQMC approach for the real-time nonequilibrium dynamics in the spin\textendash boson
model. A rather extensive swath of the full parameter space has been
explored and a detailed discussion of the convergence properties of
both the SBCI and DCCI has been made. We have compared these inchworm
expansions to several prominent, numerically exact schemes such as
QUAPI, HEOM, and MCTDH.

In general, we find that at least one of the inchworm expansions appears
to converge to the exact result in essentially all tested regions
of parameter space. This appears to include regions of parameter space
that are difficult for the QUAPI and HEOM methods. On the other hand,
at this stage the QUAPI and HEOM algorithms are simpler to employ.
In particular, more work needs to be done to fully understand the
factors that govern error growth and convergence of the various inchworm
approaches so that a general ``black-box'' implementation may be
developed which would render inchworm Monte Carlo as user-friendly
as these approaches.

In our view, the MCTDH approach is the most reliable and stable approach
for the description of dynamics in the standard spin\textendash boson
problem. The inchworm approaches presented here provide results that
appear compatible, but not quite as robust, as those produced by MCTDH.
Inchworm Monte Carlo is essentially an efficient means to stochastically
sample an exact perturbation expansion. This gives hope that the approach
may provide a general utility beyond the simplest incarnation of the
spin\textendash boson model, in cases where other methods may not
be viable. Indeed, inchworm works very well for the Anderson impurity
model, where QUAPI appears to suffer from memory length issues\cite{Segal2010Numericallyexactpath}
and MCTDH appears to have trouble in strongly correlated regimes.\cite{Wang2013Numericallyexacttime}

The biggest potential niche for the suite of inchworm Monte Carlo
approaches outlined here appears to be in a nonequilibrium setting,
such as where transport occurs between two or more reservoirs. In
such situations, MCTDH is significantly more expensive, while diagrammatic
Monte Carlo actually becomes easier to converge.\cite{Muehlbacher2008Realtimepath,Gull2010Boldlinediagrammatic,Cohen2013Numericallyexactlong}
A particularly interesting case is nonequilibrium heat transport in
the multi-bath spin\textendash boson problem.\cite{Segal2005SpinBosonThermal,Nicolin2011Nonequilibriumspin,Saito2013KondoSignaturein,Velizhanin2008Heattransportthrough,Velizhanin2010MeirWingreenformulaheat}
Here, as far as we know, only one numerically exact calculation has
been performed,\cite{Velizhanin2008Heattransportthrough,Velizhanin2010MeirWingreenformulaheat}
but owing to the numerical difficulty of the problem, a systematic
study could not be performed. This is just one example of a class
of physically important problems that may be probed in far greater
detail by the inchworm Monte Carlo methods of this work.
\begin{acknowledgments}
We would like to thank Eran Rabani and Andrés Montoya-Castillo for
discussions. GC and DRR would like to thank Andrew J. Millis and Emanuel
Gull for previous collaboration on the inchworm algorithm. DRR acknowledges
support from NSF CHE-1464802. GC and HTC acknowledge support from
the Raymond and Beverly Sackler Center for Computational Molecular
and Materials Science, Tel Aviv University.
\end{acknowledgments}

\bibliographystyle{aipnum4-1}
\bibliography{../bibtex/dqmc_spin_boson}

\begin{thebibliography}{63}%
\makeatletter
\providecommand \@ifxundefined [1]{%
 \@ifx{#1\undefined}
}%
\providecommand \@ifnum [1]{%
 \ifnum #1\expandafter \@firstoftwo
 \else \expandafter \@secondoftwo
 \fi
}%
\providecommand \@ifx [1]{%
 \ifx #1\expandafter \@firstoftwo
 \else \expandafter \@secondoftwo
 \fi
}%
\providecommand \natexlab [1]{#1}%
\providecommand \enquote  [1]{``#1''}%
\providecommand \bibnamefont  [1]{#1}%
\providecommand \bibfnamefont [1]{#1}%
\providecommand \citenamefont [1]{#1}%
\providecommand \href@noop [0]{\@secondoftwo}%
\providecommand \href [0]{\begingroup \@sanitize@url \@href}%
\providecommand \@href[1]{\@@startlink{#1}\@@href}%
\providecommand \@@href[1]{\endgroup#1\@@endlink}%
\providecommand \@sanitize@url [0]{\catcode `\\12\catcode `\$12\catcode
  `\&12\catcode `\#12\catcode `\^12\catcode `\_12\catcode `\%12\relax}%
\providecommand \@@startlink[1]{}%
\providecommand \@@endlink[0]{}%
\providecommand \url  [0]{\begingroup\@sanitize@url \@url }%
\providecommand \@url [1]{\endgroup\@href {#1}{\urlprefix }}%
\providecommand \urlprefix  [0]{URL }%
\providecommand \Eprint [0]{\href }%
\providecommand \doibase [0]{http://dx.doi.org/}%
\providecommand \selectlanguage [0]{\@gobble}%
\providecommand \bibinfo  [0]{\@secondoftwo}%
\providecommand \bibfield  [0]{\@secondoftwo}%
\providecommand \translation [1]{[#1]}%
\providecommand \BibitemOpen [0]{}%
\providecommand \bibitemStop [0]{}%
\providecommand \bibitemNoStop [0]{.\EOS\space}%
\providecommand \EOS [0]{\spacefactor3000\relax}%
\providecommand \BibitemShut  [1]{\csname bibitem#1\endcsname}%
\let\auto@bib@innerbib\@empty
\bibitem [{\citenamefont {Leggett}\ \emph {et~al.}(1987)\citenamefont
  {Leggett}, \citenamefont {Chakravarty}, \citenamefont {Dorsey}, \citenamefont
  {Fisher}, \citenamefont {Garg},\ and\ \citenamefont
  {Zwerger}}]{Leggett1987Dynamicsdissipativetwo}%
  \BibitemOpen
  \bibfield  {author} {\bibinfo {author} {\bibfnamefont {A.~J.}\ \bibnamefont
  {Leggett}}, \bibinfo {author} {\bibfnamefont {S.}~\bibnamefont
  {Chakravarty}}, \bibinfo {author} {\bibfnamefont {A.~T.~A.}\ \bibnamefont
  {Dorsey}}, \bibinfo {author} {\bibfnamefont {M.~P.~A.}\ \bibnamefont
  {Fisher}}, \bibinfo {author} {\bibfnamefont {A.}~\bibnamefont {Garg}}, \ and\
  \bibinfo {author} {\bibfnamefont {W.}~\bibnamefont {Zwerger}},\ }\href
  {\doibase 10.1103/RevModPhys.59.1} {\bibfield  {journal} {\bibinfo  {journal}
  {Rev. Mod. Phys.}\ }\textbf {\bibinfo {volume} {59}},\ \bibinfo {pages} {1}
  (\bibinfo {year} {1987})}\BibitemShut {NoStop}%
\bibitem [{\citenamefont {Weiss}(1999)}]{Weiss1999Quantumdissipativesystems}%
  \BibitemOpen
  \bibfield  {author} {\bibinfo {author} {\bibfnamefont {U.}~\bibnamefont
  {Weiss}},\ }\href@noop {} {\emph {\bibinfo {title} {{Quantum dissipative
  systems}}}},\ Vol.~\bibinfo {volume} {10}\ (\bibinfo  {publisher} {World
  Scientific Publishing Company Incorporated},\ \bibinfo {year}
  {1999})\BibitemShut {NoStop}%
\bibitem [{\citenamefont {Bader}, \citenamefont {Kuharski},\ and\ \citenamefont
  {Chandler}(1990)}]{Bader1990Rolenucleartunneling}%
  \BibitemOpen
  \bibfield  {author} {\bibinfo {author} {\bibfnamefont {J.~S.}\ \bibnamefont
  {Bader}}, \bibinfo {author} {\bibfnamefont {R.~A.}\ \bibnamefont {Kuharski}},
  \ and\ \bibinfo {author} {\bibfnamefont {D.}~\bibnamefont {Chandler}},\
  }\href
  {http://scitation.aip.org/content/aip/journal/jcp/93/1/10.1063/1.459596}
  {\bibfield  {journal} {\bibinfo  {journal} {The Journal of Chemical Physics}\
  }\textbf {\bibinfo {volume} {93}},\ \bibinfo {pages} {230} (\bibinfo {year}
  {1990})}\BibitemShut {NoStop}%
\bibitem [{\citenamefont {Warshel}, \citenamefont {Chu},\ and\ \citenamefont
  {Parson}(1989)}]{Warshel1989Dispersedpolaronsimulations}%
  \BibitemOpen
  \bibfield  {author} {\bibinfo {author} {\bibfnamefont {A.}~\bibnamefont
  {Warshel}}, \bibinfo {author} {\bibfnamefont {Z.~T.}\ \bibnamefont {Chu}}, \
  and\ \bibinfo {author} {\bibfnamefont {W.~W.}\ \bibnamefont {Parson}},\
  }\href {\doibase 10.1126/science.2675313} {\bibfield  {journal} {\bibinfo
  {journal} {Science}\ }\textbf {\bibinfo {volume} {246}},\ \bibinfo {pages}
  {112} (\bibinfo {year} {1989})}\BibitemShut {NoStop}%
\bibitem [{\citenamefont {Warshel}\ and\ \citenamefont
  {Hwang}(1986)}]{Warshel1986Simulationdynamicselectron}%
  \BibitemOpen
  \bibfield  {author} {\bibinfo {author} {\bibfnamefont {A.}~\bibnamefont
  {Warshel}}\ and\ \bibinfo {author} {\bibfnamefont {J.-K.}\ \bibnamefont
  {Hwang}},\ }\href
  {http://scitation.aip.org/content/aip/journal/jcp/84/9/10.1063/1.449981}
  {\bibfield  {journal} {\bibinfo  {journal} {The Journal of chemical physics}\
  }\textbf {\bibinfo {volume} {84}},\ \bibinfo {pages} {4938} (\bibinfo {year}
  {1986})}\BibitemShut {NoStop}%
\bibitem [{\citenamefont {Makri}(1999)}]{Makri1999LinearResponseApproximation}%
  \BibitemOpen
  \bibfield  {author} {\bibinfo {author} {\bibfnamefont {N.}~\bibnamefont
  {Makri}},\ }\href {\doibase 10.1021/jp9847540} {\bibfield  {journal}
  {\bibinfo  {journal} {The Journal of Physical Chemistry B}\ }\textbf
  {\bibinfo {volume} {103}},\ \bibinfo {pages} {2823} (\bibinfo {year}
  {1999})}\BibitemShut {NoStop}%
\bibitem [{\citenamefont {Garg}, \citenamefont {Onuchic},\ and\ \citenamefont
  {Ambegaokar}(1985)}]{Garg1985Effectfrictionelectron}%
  \BibitemOpen
  \bibfield  {author} {\bibinfo {author} {\bibfnamefont {A.}~\bibnamefont
  {Garg}}, \bibinfo {author} {\bibfnamefont {J.~N.}\ \bibnamefont {Onuchic}}, \
  and\ \bibinfo {author} {\bibfnamefont {V.}~\bibnamefont {Ambegaokar}},\
  }\href {\doibase 10.1063/1.449017} {\bibfield  {journal} {\bibinfo  {journal}
  {J. Chem. Phys.}\ }\textbf {\bibinfo {volume} {83}},\ \bibinfo {pages} {4491}
  (\bibinfo {year} {1985})}\BibitemShut {NoStop}%
\bibitem [{\citenamefont {Georgievskii}, \citenamefont {Hsu},\ and\
  \citenamefont {Marcus}(1999)}]{Georgievskii1999Linearresponsein}%
  \BibitemOpen
  \bibfield  {author} {\bibinfo {author} {\bibfnamefont {Y.}~\bibnamefont
  {Georgievskii}}, \bibinfo {author} {\bibfnamefont {C.-P.}\ \bibnamefont
  {Hsu}}, \ and\ \bibinfo {author} {\bibfnamefont {R.~A.}\ \bibnamefont
  {Marcus}},\ }\href {\doibase 10.1063/1.478425} {\bibfield  {journal}
  {\bibinfo  {journal} {J. Chem. Phys.}\ }\textbf {\bibinfo {volume} {110}},\
  \bibinfo {pages} {5307} (\bibinfo {year} {1999})}\BibitemShut {NoStop}%
\bibitem [{\citenamefont {Adolphs}\ and\ \citenamefont
  {Renger}(2006)}]{Adolphs2006HowProteinsTrigger}%
  \BibitemOpen
  \bibfield  {author} {\bibinfo {author} {\bibfnamefont {J.}~\bibnamefont
  {Adolphs}}\ and\ \bibinfo {author} {\bibfnamefont {T.}~\bibnamefont
  {Renger}},\ }\href {\doibase 10.1529/biophysj.105.079483} {\bibfield
  {journal} {\bibinfo  {journal} {Biophys. J.}\ }\textbf {\bibinfo {volume}
  {91}},\ \bibinfo {pages} {2778} (\bibinfo {year} {2006})}\BibitemShut
  {NoStop}%
\bibitem [{\citenamefont {Engel}\ \emph {et~al.}(2007)\citenamefont {Engel},
  \citenamefont {Calhoun}, \citenamefont {Read}, \citenamefont {Ahn},
  \citenamefont {Mancal}, \citenamefont {Cheng}, \citenamefont {Blankenship},\
  and\ \citenamefont {Fleming}}]{Engel2007Evidencewavelikeenergy}%
  \BibitemOpen
  \bibfield  {author} {\bibinfo {author} {\bibfnamefont {G.~S.}\ \bibnamefont
  {Engel}}, \bibinfo {author} {\bibfnamefont {T.~R.}\ \bibnamefont {Calhoun}},
  \bibinfo {author} {\bibfnamefont {E.~L.}\ \bibnamefont {Read}}, \bibinfo
  {author} {\bibfnamefont {T.~K.}\ \bibnamefont {Ahn}}, \bibinfo {author}
  {\bibfnamefont {T.}~\bibnamefont {Mancal}}, \bibinfo {author} {\bibfnamefont
  {Y.~C.}\ \bibnamefont {Cheng}}, \bibinfo {author} {\bibfnamefont {R.~E.}\
  \bibnamefont {Blankenship}}, \ and\ \bibinfo {author} {\bibfnamefont {G.~R.}\
  \bibnamefont {Fleming}},\ }\href {\doibase 10.1038/nature05678} {\bibfield
  {journal} {\bibinfo  {journal} {Nature}\ }\textbf {\bibinfo {volume} {446}},\
  \bibinfo {pages} {782} (\bibinfo {year} {2007})}\BibitemShut {NoStop}%
\bibitem [{\citenamefont {Lee}, \citenamefont {Cheng},\ and\ \citenamefont
  {Fleming}(2007)}]{Lee2007Coherencedynamicsin}%
  \BibitemOpen
  \bibfield  {author} {\bibinfo {author} {\bibfnamefont {H.}~\bibnamefont
  {Lee}}, \bibinfo {author} {\bibfnamefont {Y.~C.}\ \bibnamefont {Cheng}}, \
  and\ \bibinfo {author} {\bibfnamefont {G.~R.}\ \bibnamefont {Fleming}},\
  }\href {\doibase 10.1126/science.1142188} {\bibfield  {journal} {\bibinfo
  {journal} {Science}\ }\textbf {\bibinfo {volume} {316}},\ \bibinfo {pages}
  {1462} (\bibinfo {year} {2007})}\BibitemShut {NoStop}%
\bibitem [{\citenamefont {Panitchayangkoon}\ \emph {et~al.}(2010)\citenamefont
  {Panitchayangkoon}, \citenamefont {Hayes}, \citenamefont {Fransted},
  \citenamefont {Caram}, \citenamefont {Harel}, \citenamefont {Wen},
  \citenamefont {Blankenship},\ and\ \citenamefont
  {Engel}}]{Panitchayangkoon2010Longlivedquantum}%
  \BibitemOpen
  \bibfield  {author} {\bibinfo {author} {\bibfnamefont {G.}~\bibnamefont
  {Panitchayangkoon}}, \bibinfo {author} {\bibfnamefont {D.}~\bibnamefont
  {Hayes}}, \bibinfo {author} {\bibfnamefont {K.~A.}\ \bibnamefont {Fransted}},
  \bibinfo {author} {\bibfnamefont {J.~R.}\ \bibnamefont {Caram}}, \bibinfo
  {author} {\bibfnamefont {E.}~\bibnamefont {Harel}}, \bibinfo {author}
  {\bibfnamefont {J.}~\bibnamefont {Wen}}, \bibinfo {author} {\bibfnamefont
  {R.~E.}\ \bibnamefont {Blankenship}}, \ and\ \bibinfo {author} {\bibfnamefont
  {G.~S.}\ \bibnamefont {Engel}},\ }\href
  {http://www.pubmedcentral.nih.gov/articlerender.fcgi?artid=2919932{\&}tool=pmcentrez{\&}rendertype=abstract}
  {\bibfield  {journal} {\bibinfo  {journal} {Proc. Natl. Acad. Sci. U. S. A.}\
  }\textbf {\bibinfo {volume} {107}},\ \bibinfo {pages} {12766} (\bibinfo
  {year} {2010})},\ \Eprint {http://arxiv.org/abs/1001.5108} {1001.5108}
  \BibitemShut {NoStop}%
\bibitem [{\citenamefont {Collini}\ and\ \citenamefont
  {Scholes}(2009)}]{Collini2009CoherentIntrachainEnergy}%
  \BibitemOpen
  \bibfield  {author} {\bibinfo {author} {\bibfnamefont {E.}~\bibnamefont
  {Collini}}\ and\ \bibinfo {author} {\bibfnamefont {G.~D.}\ \bibnamefont
  {Scholes}},\ }\href {http://www.sciencemag.org/content/323/5912/369.short}
  {\bibfield  {journal} {\bibinfo  {journal} {Science}\ }\textbf {\bibinfo
  {volume} {323}},\ \bibinfo {pages} {369} (\bibinfo {year}
  {2009})}\BibitemShut {NoStop}%
\bibitem [{\citenamefont {Br{\'{e}}das}\ and\ \citenamefont
  {Silbey}(2009)}]{Bredas2009ExcitonsSurfAlong}%
  \BibitemOpen
  \bibfield  {author} {\bibinfo {author} {\bibfnamefont {J.~L.}\ \bibnamefont
  {Br{\'{e}}das}}\ and\ \bibinfo {author} {\bibfnamefont {R.}~\bibnamefont
  {Silbey}},\ }\href@noop {} {\bibfield  {journal} {\bibinfo  {journal}
  {Science}\ }\textbf {\bibinfo {volume} {323}},\ \bibinfo {pages} {348}
  (\bibinfo {year} {2009})}\BibitemShut {NoStop}%
\bibitem [{\citenamefont {Fisher}\ and\ \citenamefont
  {Zwerger}(1985)}]{Fisher1985QuantumBrownianmotion}%
  \BibitemOpen
  \bibfield  {author} {\bibinfo {author} {\bibfnamefont {M.~P.~A.}\
  \bibnamefont {Fisher}}\ and\ \bibinfo {author} {\bibfnamefont
  {W.}~\bibnamefont {Zwerger}},\ }\href {\doibase 10.1103/PhysRevB.32.6190}
  {\bibfield  {journal} {\bibinfo  {journal} {Physical Review B}\ }\textbf
  {\bibinfo {volume} {32}},\ \bibinfo {pages} {6190} (\bibinfo {year}
  {1985})}\BibitemShut {NoStop}%
\bibitem [{\citenamefont {Prokof'ev}\ and\ \citenamefont
  {Stamp}(1996)}]{ProkofEv1996Quantumrelaxationmagnetisation}%
  \BibitemOpen
  \bibfield  {author} {\bibinfo {author} {\bibfnamefont {N.~V.}\ \bibnamefont
  {Prokof'ev}}\ and\ \bibinfo {author} {\bibfnamefont {P.~C.~E.}\ \bibnamefont
  {Stamp}},\ }\href {http://link.springer.com/article/10.1007/BF00754094}
  {\bibfield  {journal} {\bibinfo  {journal} {J. Low Temp. Phys.}\ }\textbf
  {\bibinfo {volume} {104}},\ \bibinfo {pages} {143} (\bibinfo {year}
  {1996})}\BibitemShut {NoStop}%
\bibitem [{\citenamefont {Prokof'ev}\ and\ \citenamefont
  {Stamp}(1998)}]{ProkofEv1998Lowtemperaturequantum}%
  \BibitemOpen
  \bibfield  {author} {\bibinfo {author} {\bibfnamefont {N.~V.}\ \bibnamefont
  {Prokof'ev}}\ and\ \bibinfo {author} {\bibfnamefont {P.~C.~E.}\ \bibnamefont
  {Stamp}},\ }\href
  {http://journals.aps.org/prl/abstract/10.1103/PhysRevLett.80.5794} {\bibfield
   {journal} {\bibinfo  {journal} {Phys. Rev. Lett.}\ }\textbf {\bibinfo
  {volume} {80}},\ \bibinfo {pages} {5794} (\bibinfo {year}
  {1998})}\BibitemShut {NoStop}%
\bibitem [{\citenamefont {Goldstein}\ \emph {et~al.}(2013)\citenamefont
  {Goldstein}, \citenamefont {Devoret}, \citenamefont {Houzet},\ and\
  \citenamefont {Glazman}}]{Goldstein2013InelasticMicrowavePhoton}%
  \BibitemOpen
  \bibfield  {author} {\bibinfo {author} {\bibfnamefont {M.}~\bibnamefont
  {Goldstein}}, \bibinfo {author} {\bibfnamefont {M.~H.}\ \bibnamefont
  {Devoret}}, \bibinfo {author} {\bibfnamefont {M.}~\bibnamefont {Houzet}}, \
  and\ \bibinfo {author} {\bibfnamefont {L.~I.}\ \bibnamefont {Glazman}},\
  }\href {\doibase 10.1103/PhysRevLett.110.017002} {\bibfield  {journal}
  {\bibinfo  {journal} {Phys. Rev. Lett.}\ }\textbf {\bibinfo {volume} {110}},\
  \bibinfo {pages} {017002} (\bibinfo {year} {2013})}\BibitemShut {NoStop}%
\bibitem [{\citenamefont {Fink}\ \emph {et~al.}(2008)\citenamefont {Fink},
  \citenamefont {G{\"o}ppl}, \citenamefont {Baur}, \citenamefont {Bianchetti},
  \citenamefont {Leek}, \citenamefont {Blais},\ and\ \citenamefont
  {Wallraff}}]{Fink2008ClimbingJaynesCummingsladder}%
  \BibitemOpen
  \bibfield  {author} {\bibinfo {author} {\bibfnamefont {J.~M.}\ \bibnamefont
  {Fink}}, \bibinfo {author} {\bibfnamefont {M.}~\bibnamefont {G{\"o}ppl}},
  \bibinfo {author} {\bibfnamefont {M.}~\bibnamefont {Baur}}, \bibinfo {author}
  {\bibfnamefont {R.}~\bibnamefont {Bianchetti}}, \bibinfo {author}
  {\bibfnamefont {P.~J.}\ \bibnamefont {Leek}}, \bibinfo {author}
  {\bibfnamefont {A.}~\bibnamefont {Blais}}, \ and\ \bibinfo {author}
  {\bibfnamefont {A.}~\bibnamefont {Wallraff}},\ }\href {\doibase
  10.1038/nature07112} {\bibfield  {journal} {\bibinfo  {journal} {Nature}\
  }\textbf {\bibinfo {volume} {454}},\ \bibinfo {pages} {315} (\bibinfo {year}
  {2008})}\BibitemShut {NoStop}%
\bibitem [{\citenamefont {Solano}, \citenamefont {Agarwal},\ and\ \citenamefont
  {Walther}(2003)}]{Solano2003StrongDrivingAssisted}%
  \BibitemOpen
  \bibfield  {author} {\bibinfo {author} {\bibfnamefont {E.}~\bibnamefont
  {Solano}}, \bibinfo {author} {\bibfnamefont {G.~S.}\ \bibnamefont {Agarwal}},
  \ and\ \bibinfo {author} {\bibfnamefont {H.}~\bibnamefont {Walther}},\ }\href
  {\doibase 10.1103/PhysRevLett.90.027903} {\bibfield  {journal} {\bibinfo
  {journal} {Phys. Rev. Lett.}\ }\textbf {\bibinfo {volume} {90}},\ \bibinfo
  {pages} {027903} (\bibinfo {year} {2003})}\BibitemShut {NoStop}%
\bibitem [{\citenamefont {Casanova}\ \emph {et~al.}(2010)\citenamefont
  {Casanova}, \citenamefont {Romero}, \citenamefont {Lizuain}, \citenamefont
  {Garc\'{i}a-Ripoll},\ and\ \citenamefont
  {Solano}}]{Casanova2010DeepStrongCoupling}%
  \BibitemOpen
  \bibfield  {author} {\bibinfo {author} {\bibfnamefont {J.}~\bibnamefont
  {Casanova}}, \bibinfo {author} {\bibfnamefont {G.}~\bibnamefont {Romero}},
  \bibinfo {author} {\bibfnamefont {I.}~\bibnamefont {Lizuain}}, \bibinfo
  {author} {\bibfnamefont {J.~J.}\ \bibnamefont {Garc\'{i}a-Ripoll}}, \ and\
  \bibinfo {author} {\bibfnamefont {E.}~\bibnamefont {Solano}},\ }\href
  {\doibase 10.1103/PhysRevLett.105.263603} {\bibfield  {journal} {\bibinfo
  {journal} {Phys. Rev. Lett.}\ }\textbf {\bibinfo {volume} {105}},\ \bibinfo
  {pages} {263603} (\bibinfo {year} {2010})}\BibitemShut {NoStop}%
\bibitem [{\citenamefont {Makarov}\ and\ \citenamefont
  {Makri}(1994)}]{Makarov1994Pathintegralsdissipative}%
  \BibitemOpen
  \bibfield  {author} {\bibinfo {author} {\bibfnamefont {D.~E.}\ \bibnamefont
  {Makarov}}\ and\ \bibinfo {author} {\bibfnamefont {N.}~\bibnamefont
  {Makri}},\ }\href {\doibase 10.1016/0009-2614(94)00275-4} {\bibfield
  {journal} {\bibinfo  {journal} {Chem. Phys. Lett.}\ }\textbf {\bibinfo
  {volume} {221}},\ \bibinfo {pages} {482} (\bibinfo {year}
  {1994})}\BibitemShut {NoStop}%
\bibitem [{\citenamefont {Makri}(1995)}]{Makri1995Numericalpathintegral}%
  \BibitemOpen
  \bibfield  {author} {\bibinfo {author} {\bibfnamefont {N.}~\bibnamefont
  {Makri}},\ }\href {\doibase 10.1063/1.531046} {\bibfield  {journal} {\bibinfo
   {journal} {J. Math. Phys.}\ }\textbf {\bibinfo {volume} {36}},\ \bibinfo
  {pages} {2430} (\bibinfo {year} {1995})}\BibitemShut {NoStop}%
\bibitem [{\citenamefont {Makri}\ and\ \citenamefont
  {Makarov}(1995)}]{Makri1995Tensorpropagatoriterative}%
  \BibitemOpen
  \bibfield  {author} {\bibinfo {author} {\bibfnamefont {N.}~\bibnamefont
  {Makri}}\ and\ \bibinfo {author} {\bibfnamefont {D.~E.}\ \bibnamefont
  {Makarov}},\ }\href {\doibase 10.1063/1.469509} {\bibfield  {journal}
  {\bibinfo  {journal} {J. Chem. Phys.}\ }\textbf {\bibinfo {volume} {102}},\
  \bibinfo {pages} {4600} (\bibinfo {year} {1995})}\BibitemShut {NoStop}%
\bibitem [{\citenamefont {Makri}\ \emph {et~al.}(1996)\citenamefont {Makri},
  \citenamefont {Sim}, \citenamefont {Makarov},\ and\ \citenamefont
  {Topaler}}]{Makri1996Longtimequantum}%
  \BibitemOpen
  \bibfield  {author} {\bibinfo {author} {\bibfnamefont {N.}~\bibnamefont
  {Makri}}, \bibinfo {author} {\bibfnamefont {E.}~\bibnamefont {Sim}}, \bibinfo
  {author} {\bibfnamefont {D.~E.}\ \bibnamefont {Makarov}}, \ and\ \bibinfo
  {author} {\bibfnamefont {M.}~\bibnamefont {Topaler}},\ }\href
  {http://www.pnas.org/content/93/9/3926.short} {\bibfield  {journal} {\bibinfo
   {journal} {Proc. Natl. Acad. Sci. U. S. A.}\ }\textbf {\bibinfo {volume}
  {93}},\ \bibinfo {pages} {3926} (\bibinfo {year} {1996})}\BibitemShut
  {NoStop}%
\bibitem [{\citenamefont {Tanimura}\ and\ \citenamefont
  {Kubo}(1989)}]{Tanimura1989TimeEvolutionQuantum}%
  \BibitemOpen
  \bibfield  {author} {\bibinfo {author} {\bibfnamefont {Y.}~\bibnamefont
  {Tanimura}}\ and\ \bibinfo {author} {\bibfnamefont {R.}~\bibnamefont
  {Kubo}},\ }\href {\doibase 10.1143/JPSJ.58.101} {\bibfield  {journal}
  {\bibinfo  {journal} {Journal of the Physical Society of Japan}\ }\textbf
  {\bibinfo {volume} {58}},\ \bibinfo {pages} {101} (\bibinfo {year}
  {1989})}\BibitemShut {NoStop}%
\bibitem [{\citenamefont {Ishizaki}\ and\ \citenamefont
  {Fleming}(2009)}]{Ishizaki2009Unifiedtreatmentquantum}%
  \BibitemOpen
  \bibfield  {author} {\bibinfo {author} {\bibfnamefont {A.}~\bibnamefont
  {Ishizaki}}\ and\ \bibinfo {author} {\bibfnamefont {G.~R.}\ \bibnamefont
  {Fleming}},\ }\href {\doibase 10.1063/1.3155372} {\bibfield  {journal}
  {\bibinfo  {journal} {J. Chem. Phys.}\ }\textbf {\bibinfo {volume} {130}},\
  \bibinfo {pages} {234111} (\bibinfo {year} {2009})}\BibitemShut {NoStop}%
\bibitem [{\citenamefont {Str{\"{u}}mpfer}\ and\ \citenamefont
  {Schulten}(2012)}]{Struempfer2012OpenQuantumDynamics}%
  \BibitemOpen
  \bibfield  {author} {\bibinfo {author} {\bibfnamefont {J.}~\bibnamefont
  {Str{\"{u}}mpfer}}\ and\ \bibinfo {author} {\bibfnamefont {K.}~\bibnamefont
  {Schulten}},\ }\href {\doibase 10.1021/ct3003833} {\bibfield  {journal}
  {\bibinfo  {journal} {J. Chem. Theory Comput.}\ }\textbf {\bibinfo {volume}
  {8}},\ \bibinfo {pages} {2808} (\bibinfo {year} {2012})}\BibitemShut
  {NoStop}%
\bibitem [{\citenamefont {Thoss}, \citenamefont {Wang},\ and\ \citenamefont
  {Miller}(2001)}]{Thoss2001Selfconsistenthybrid}%
  \BibitemOpen
  \bibfield  {author} {\bibinfo {author} {\bibfnamefont {M.}~\bibnamefont
  {Thoss}}, \bibinfo {author} {\bibfnamefont {H.}~\bibnamefont {Wang}}, \ and\
  \bibinfo {author} {\bibfnamefont {W.~H.}\ \bibnamefont {Miller}},\ }\href
  {\doibase 10.1063/1.1385562} {\bibfield  {journal} {\bibinfo  {journal} {J.
  Chem. Phys.}\ }\textbf {\bibinfo {volume} {115}},\ \bibinfo {pages} {2991}
  (\bibinfo {year} {2001})}\BibitemShut {NoStop}%
\bibitem [{\citenamefont {Wang}, \citenamefont {Thoss},\ and\ \citenamefont
  {Miller}(2001)}]{Wang2001Systematicconvergencein}%
  \BibitemOpen
  \bibfield  {author} {\bibinfo {author} {\bibfnamefont {H.}~\bibnamefont
  {Wang}}, \bibinfo {author} {\bibfnamefont {M.}~\bibnamefont {Thoss}}, \ and\
  \bibinfo {author} {\bibfnamefont {W.~H.}\ \bibnamefont {Miller}},\ }\href
  {\doibase 10.1063/1.1385561} {\bibfield  {journal} {\bibinfo  {journal} {J.
  Chem. Phys.}\ }\textbf {\bibinfo {volume} {115}},\ \bibinfo {pages} {2979}
  (\bibinfo {year} {2001})}\BibitemShut {NoStop}%
\bibitem [{\citenamefont {Wang}\ and\ \citenamefont
  {Thoss}(2003)}]{Wang2003Multilayerformulationmulticonfiguration}%
  \BibitemOpen
  \bibfield  {author} {\bibinfo {author} {\bibfnamefont {H.}~\bibnamefont
  {Wang}}\ and\ \bibinfo {author} {\bibfnamefont {M.}~\bibnamefont {Thoss}},\
  }\href {\doibase 10.1063/1.1580111} {\bibfield  {journal} {\bibinfo
  {journal} {J. Chem. Phys.}\ }\textbf {\bibinfo {volume} {119}},\ \bibinfo
  {pages} {1289} (\bibinfo {year} {2003})}\BibitemShut {NoStop}%
\bibitem [{\citenamefont {Mak}\ and\ \citenamefont
  {Chandler}(1990)}]{Mak1990Solvingsignproblem}%
  \BibitemOpen
  \bibfield  {author} {\bibinfo {author} {\bibfnamefont {C.~H.}\ \bibnamefont
  {Mak}}\ and\ \bibinfo {author} {\bibfnamefont {D.}~\bibnamefont {Chandler}},\
  }\href {\doibase 10.1103/PhysRevA.41.5709} {\bibfield  {journal} {\bibinfo
  {journal} {Phys. Rev. A}\ }\textbf {\bibinfo {volume} {41}},\ \bibinfo
  {pages} {5709} (\bibinfo {year} {1990})}\BibitemShut {NoStop}%
\bibitem [{\citenamefont {Mak}\ and\ \citenamefont
  {Chandler}(1991)}]{Mak1991Coherentincoherenttransition}%
  \BibitemOpen
  \bibfield  {author} {\bibinfo {author} {\bibfnamefont {C.~H.}\ \bibnamefont
  {Mak}}\ and\ \bibinfo {author} {\bibfnamefont {D.}~\bibnamefont {Chandler}},\
  }\href {\doibase 10.1103/PhysRevA.44.2352} {\bibfield  {journal} {\bibinfo
  {journal} {Physical Review A}\ }\textbf {\bibinfo {volume} {44}},\ \bibinfo
  {pages} {2352} (\bibinfo {year} {1991})}\BibitemShut {NoStop}%
\bibitem [{\citenamefont {Egger}\ and\ \citenamefont
  {Weiss}(1992)}]{Egger1992QuantumMonteCarlo}%
  \BibitemOpen
  \bibfield  {author} {\bibinfo {author} {\bibfnamefont {R.}~\bibnamefont
  {Egger}}\ and\ \bibinfo {author} {\bibfnamefont {U.}~\bibnamefont {Weiss}},\
  }\href {\doibase 10.1007/BF01320834} {\bibfield  {journal} {\bibinfo
  {journal} {Zeitschrift f{\"{u}}r Phys. B Condens. Matter}\ }\textbf {\bibinfo
  {volume} {89}},\ \bibinfo {pages} {97} (\bibinfo {year} {1992})}\BibitemShut
  {NoStop}%
\bibitem [{\citenamefont {Egger}\ and\ \citenamefont
  {Mak}(1994)}]{Egger1994Lowtemperaturedynamical}%
  \BibitemOpen
  \bibfield  {author} {\bibinfo {author} {\bibfnamefont {R.}~\bibnamefont
  {Egger}}\ and\ \bibinfo {author} {\bibfnamefont {C.~H.}\ \bibnamefont
  {Mak}},\ }\href {\doibase 10.1103/PhysRevB.50.15210} {\bibfield  {journal}
  {\bibinfo  {journal} {Phys. Rev. B}\ }\textbf {\bibinfo {volume} {50}},\
  \bibinfo {pages} {210} (\bibinfo {year} {1994})}\BibitemShut {NoStop}%
\bibitem [{\citenamefont {Egger}, \citenamefont {M{\"{u}}hlbacher},\ and\
  \citenamefont {Mak}(2000)}]{Egger2000PathintegralMonte}%
  \BibitemOpen
  \bibfield  {author} {\bibinfo {author} {\bibfnamefont {R.}~\bibnamefont
  {Egger}}, \bibinfo {author} {\bibfnamefont {L.}~\bibnamefont
  {M{\"{u}}hlbacher}}, \ and\ \bibinfo {author} {\bibfnamefont {C.~H.}\
  \bibnamefont {Mak}},\ }\href {\doibase 10.1103/PhysRevE.61.5961} {\bibfield
  {journal} {\bibinfo  {journal} {Phys. Rev. E}\ }\textbf {\bibinfo {volume}
  {61}},\ \bibinfo {pages} {5961} (\bibinfo {year} {2000})}\BibitemShut
  {NoStop}%
\bibitem [{\citenamefont {Prokof'ev}\ and\ \citenamefont
  {Svistunov}(1998)}]{Prokofev1998PolaronProblemby}%
  \BibitemOpen
  \bibfield  {author} {\bibinfo {author} {\bibfnamefont {N.~V.}\ \bibnamefont
  {Prokof'ev}}\ and\ \bibinfo {author} {\bibfnamefont {B.~V.}\ \bibnamefont
  {Svistunov}},\ }\href {\doibase 10.1103/PhysRevLett.81.2514} {\bibfield
  {journal} {\bibinfo  {journal} {Physical Review Letters}\ }\textbf {\bibinfo
  {volume} {81}},\ \bibinfo {pages} {2514} (\bibinfo {year}
  {1998})}\BibitemShut {NoStop}%
\bibitem [{\citenamefont {Prokof'ev}\ and\ \citenamefont
  {Svistunov}(2008{\natexlab{a}})}]{Prokofev2008BolddiagrammaticMonte}%
  \BibitemOpen
  \bibfield  {author} {\bibinfo {author} {\bibfnamefont {N.~V.}\ \bibnamefont
  {Prokof'ev}}\ and\ \bibinfo {author} {\bibfnamefont {B.~V.}\ \bibnamefont
  {Svistunov}},\ }\href {\doibase 10.1103/PhysRevB.77.125101} {\bibfield
  {journal} {\bibinfo  {journal} {Physical Review B}\ }\textbf {\bibinfo
  {volume} {77}},\ \bibinfo {pages} {125101} (\bibinfo {year}
  {2008}{\natexlab{a}})}\BibitemShut {NoStop}%
\bibitem [{\citenamefont {Prokof'ev}\ and\ \citenamefont
  {Svistunov}(2008{\natexlab{b}})}]{Prokofev2008Fermipolaronproblem}%
  \BibitemOpen
  \bibfield  {author} {\bibinfo {author} {\bibfnamefont {N.}~\bibnamefont
  {Prokof'ev}}\ and\ \bibinfo {author} {\bibfnamefont {B.}~\bibnamefont
  {Svistunov}},\ }\href {\doibase 10.1103/PhysRevB.77.020408} {\bibfield
  {journal} {\bibinfo  {journal} {Physical Review B}\ }\textbf {\bibinfo
  {volume} {77}},\ \bibinfo {pages} {020408} (\bibinfo {year}
  {2008}{\natexlab{b}})}\BibitemShut {NoStop}%
\bibitem [{\citenamefont {Van~Houcke}\ \emph {et~al.}(2010)\citenamefont
  {Van~Houcke}, \citenamefont {Kozik}, \citenamefont {Prokof'ev},\ and\
  \citenamefont {Svistunov}}]{VanHoucke2010DiagrammaticMonteCarlo}%
  \BibitemOpen
  \bibfield  {author} {\bibinfo {author} {\bibfnamefont {K.}~\bibnamefont
  {Van~Houcke}}, \bibinfo {author} {\bibfnamefont {E.}~\bibnamefont {Kozik}},
  \bibinfo {author} {\bibfnamefont {N.}~\bibnamefont {Prokof'ev}}, \ and\
  \bibinfo {author} {\bibfnamefont {B.}~\bibnamefont {Svistunov}},\ }\href
  {\doibase 10.1016/j.phpro.2010.09.034} {\bibfield  {journal} {\bibinfo
  {journal} {Physics Procedia}\ }\bibinfo {series} {Computer {Simulations}
  {Studies} in {Condensed} {Matter} {Physics} {XXI}},\ \textbf {\bibinfo
  {volume} {6}},\ \bibinfo {pages} {95} (\bibinfo {year} {2010})}\BibitemShut
  {NoStop}%
\bibitem [{\citenamefont {Gull}\ \emph {et~al.}(2011)\citenamefont {Gull},
  \citenamefont {Millis}, \citenamefont {Lichtenstein}, \citenamefont
  {Rubtsov}, \citenamefont {Troyer},\ and\ \citenamefont
  {Werner}}]{Gull2011ContinuoustimeMonte}%
  \BibitemOpen
  \bibfield  {author} {\bibinfo {author} {\bibfnamefont {E.}~\bibnamefont
  {Gull}}, \bibinfo {author} {\bibfnamefont {A.~J.}\ \bibnamefont {Millis}},
  \bibinfo {author} {\bibfnamefont {A.~I.}\ \bibnamefont {Lichtenstein}},
  \bibinfo {author} {\bibfnamefont {A.~N.}\ \bibnamefont {Rubtsov}}, \bibinfo
  {author} {\bibfnamefont {M.}~\bibnamefont {Troyer}}, \ and\ \bibinfo {author}
  {\bibfnamefont {P.}~\bibnamefont {Werner}},\ }\href@noop {} {\bibfield
  {journal} {\bibinfo  {journal} {Rev. Mod. Phys.}\ }\textbf {\bibinfo {volume}
  {83}},\ \bibinfo {pages} {349} (\bibinfo {year} {2011})},\ \Eprint
  {http://arxiv.org/abs/1012.4474} {1012.4474} \BibitemShut {NoStop}%
\bibitem [{\citenamefont {M{\"{u}}hlbacher}\ and\ \citenamefont
  {Rabani}(2008)}]{Muehlbacher2008Realtimepath}%
  \BibitemOpen
  \bibfield  {author} {\bibinfo {author} {\bibfnamefont {L.}~\bibnamefont
  {M{\"{u}}hlbacher}}\ and\ \bibinfo {author} {\bibfnamefont {E.}~\bibnamefont
  {Rabani}},\ }\href {http://link.aps.org/doi/10.1103/PhysRevLett.100.176403}
  {\bibfield  {journal} {\bibinfo  {journal} {Phys. Rev. Lett.}\ }\textbf
  {\bibinfo {volume} {100}},\ \bibinfo {pages} {176403} (\bibinfo {year}
  {2008})},\ \Eprint {http://arxiv.org/abs/0707.0956} {0707.0956} \BibitemShut
  {NoStop}%
\bibitem [{\citenamefont {Werner}, \citenamefont {Oka},\ and\ \citenamefont
  {Millis}(2009)}]{Werner2009DiagrammaticMonteCarlo}%
  \BibitemOpen
  \bibfield  {author} {\bibinfo {author} {\bibfnamefont {P.}~\bibnamefont
  {Werner}}, \bibinfo {author} {\bibfnamefont {T.}~\bibnamefont {Oka}}, \ and\
  \bibinfo {author} {\bibfnamefont {A.~J.}\ \bibnamefont {Millis}},\ }\href
  {\doibase 10.1103/PhysRevB.79.035320} {\bibfield  {journal} {\bibinfo
  {journal} {Phys. Rev. B}\ }\textbf {\bibinfo {volume} {79}},\ \bibinfo
  {pages} {035320} (\bibinfo {year} {2009})}\BibitemShut {NoStop}%
\bibitem [{\citenamefont {Cohen}\ \emph
  {et~al.}(2014{\natexlab{a}})\citenamefont {Cohen}, \citenamefont {Reichman},
  \citenamefont {Millis},\ and\ \citenamefont
  {Gull}}]{Cohen2014Greensfunctionsfroma}%
  \BibitemOpen
  \bibfield  {author} {\bibinfo {author} {\bibfnamefont {G.}~\bibnamefont
  {Cohen}}, \bibinfo {author} {\bibfnamefont {D.~R.}\ \bibnamefont {Reichman}},
  \bibinfo {author} {\bibfnamefont {A.~J.}\ \bibnamefont {Millis}}, \ and\
  \bibinfo {author} {\bibfnamefont {E.}~\bibnamefont {Gull}},\ }\href@noop {}
  {\bibfield  {journal} {\bibinfo  {journal} {Phys. Rev. B}\ }\textbf {\bibinfo
  {volume} {89}},\ \bibinfo {pages} {115139} (\bibinfo {year}
  {2014}{\natexlab{a}})}\BibitemShut {NoStop}%
\bibitem [{\citenamefont {Cohen}\ \emph
  {et~al.}(2014{\natexlab{b}})\citenamefont {Cohen}, \citenamefont {Gull},
  \citenamefont {Reichman},\ and\ \citenamefont
  {Millis}}]{Cohen2014Greensfunctionsfrom}%
  \BibitemOpen
  \bibfield  {author} {\bibinfo {author} {\bibfnamefont {G.}~\bibnamefont
  {Cohen}}, \bibinfo {author} {\bibfnamefont {E.}~\bibnamefont {Gull}},
  \bibinfo {author} {\bibfnamefont {D.~R.}\ \bibnamefont {Reichman}}, \ and\
  \bibinfo {author} {\bibfnamefont {A.~J.}\ \bibnamefont {Millis}},\
  }\href@noop {} {\bibfield  {journal} {\bibinfo  {journal} {Phys. Rev. Lett.}\
  }\textbf {\bibinfo {volume} {112}},\ \bibinfo {pages} {146802} (\bibinfo
  {year} {2014}{\natexlab{b}})}\BibitemShut {NoStop}%
\bibitem [{\citenamefont {Cohen}\ \emph {et~al.}(2013)\citenamefont {Cohen},
  \citenamefont {Gull}, \citenamefont {Reichman}, \citenamefont {Millis},\ and\
  \citenamefont {Rabani}}]{Cohen2013Numericallyexactlong}%
  \BibitemOpen
  \bibfield  {author} {\bibinfo {author} {\bibfnamefont {G.}~\bibnamefont
  {Cohen}}, \bibinfo {author} {\bibfnamefont {E.}~\bibnamefont {Gull}},
  \bibinfo {author} {\bibfnamefont {D.~R.}\ \bibnamefont {Reichman}}, \bibinfo
  {author} {\bibfnamefont {A.~J.}\ \bibnamefont {Millis}}, \ and\ \bibinfo
  {author} {\bibfnamefont {E.}~\bibnamefont {Rabani}},\ }\href@noop {}
  {\bibfield  {journal} {\bibinfo  {journal} {Phys. Rev. B}\ }\textbf {\bibinfo
  {volume} {87}},\ \bibinfo {pages} {195108} (\bibinfo {year}
  {2013})}\BibitemShut {NoStop}%
\bibitem [{\citenamefont {Cohen}\ \emph {et~al.}(2015)\citenamefont {Cohen},
  \citenamefont {Gull}, \citenamefont {Reichman},\ and\ \citenamefont
  {Millis}}]{Cohen2015Tamingdynamicalsign}%
  \BibitemOpen
  \bibfield  {author} {\bibinfo {author} {\bibfnamefont {G.}~\bibnamefont
  {Cohen}}, \bibinfo {author} {\bibfnamefont {E.}~\bibnamefont {Gull}},
  \bibinfo {author} {\bibfnamefont {D.~R.}\ \bibnamefont {Reichman}}, \ and\
  \bibinfo {author} {\bibfnamefont {A.~J.}\ \bibnamefont {Millis}},\ }\href
  {\doibase 10.1103/PhysRevLett.115.266802} {\bibfield  {journal} {\bibinfo
  {journal} {Phys. Rev. Lett.}\ }\textbf {\bibinfo {volume} {115}},\ \bibinfo
  {pages} {266802} (\bibinfo {year} {2015})}\BibitemShut {NoStop}%
\bibitem [{\citenamefont {Montoya-Castillo}, \citenamefont {Berkelbach},\ and\
  \citenamefont
  {Reichman}(2015)}]{Montoya-Castillo2015ExtendingapplicabilityRedfield}%
  \BibitemOpen
  \bibfield  {author} {\bibinfo {author} {\bibfnamefont {A.}~\bibnamefont
  {Montoya-Castillo}}, \bibinfo {author} {\bibfnamefont {T.~C.}\ \bibnamefont
  {Berkelbach}}, \ and\ \bibinfo {author} {\bibfnamefont {D.~R.}\ \bibnamefont
  {Reichman}},\ }\href {\doibase 10.1063/1.4935443} {\bibfield  {journal}
  {\bibinfo  {journal} {J. Chem. Phys.}\ }\textbf {\bibinfo {volume} {143}},\
  \bibinfo {pages} {194108} (\bibinfo {year} {2015})}\BibitemShut {NoStop}%
\bibitem [{\citenamefont {Hu}, \citenamefont {Xu},\ and\ \citenamefont
  {Yan}(2010)}]{Hu2010CommunicationPadespectrum}%
  \BibitemOpen
  \bibfield  {author} {\bibinfo {author} {\bibfnamefont {J.}~\bibnamefont
  {Hu}}, \bibinfo {author} {\bibfnamefont {R.-X.}\ \bibnamefont {Xu}}, \ and\
  \bibinfo {author} {\bibfnamefont {Y.}~\bibnamefont {Yan}},\ }\href@noop {}
  {\bibfield  {journal} {\bibinfo  {journal} {J. Chem. Phys.}\ }\textbf
  {\bibinfo {volume} {133}},\ \bibinfo {pages} {101106} (\bibinfo {year}
  {2010})}\BibitemShut {NoStop}%
\bibitem [{\citenamefont {Hu}\ \emph {et~al.}(2011)\citenamefont {Hu},
  \citenamefont {Luo}, \citenamefont {Jiang}, \citenamefont {Xu},\ and\
  \citenamefont {Yan}}]{Hu2011Padespectrumdecompositions}%
  \BibitemOpen
  \bibfield  {author} {\bibinfo {author} {\bibfnamefont {J.}~\bibnamefont
  {Hu}}, \bibinfo {author} {\bibfnamefont {M.}~\bibnamefont {Luo}}, \bibinfo
  {author} {\bibfnamefont {F.}~\bibnamefont {Jiang}}, \bibinfo {author}
  {\bibfnamefont {R.-X.}\ \bibnamefont {Xu}}, \ and\ \bibinfo {author}
  {\bibfnamefont {Y.}~\bibnamefont {Yan}},\ }\href@noop {} {\bibfield
  {journal} {\bibinfo  {journal} {J. Chem. Phys.}\ }\textbf {\bibinfo {volume}
  {134}},\ \bibinfo {pages} {244106} (\bibinfo {year} {2011})}\BibitemShut
  {NoStop}%
\bibitem [{\citenamefont {Yan}(2014)}]{Yan2014Theoryopenquantum}%
  \BibitemOpen
  \bibfield  {author} {\bibinfo {author} {\bibfnamefont {Y.}~\bibnamefont
  {Yan}},\ }\href@noop {} {\bibfield  {journal} {\bibinfo  {journal} {J. Chem.
  Phys.}\ }\textbf {\bibinfo {volume} {140}},\ \bibinfo {pages} {054105}
  (\bibinfo {year} {2014})}\BibitemShut {NoStop}%
\bibitem [{\citenamefont {Moix}\ and\ \citenamefont
  {Cao}(2013)}]{Moix2013hybridstochastichierarchy}%
  \BibitemOpen
  \bibfield  {author} {\bibinfo {author} {\bibfnamefont {J.~M.}\ \bibnamefont
  {Moix}}\ and\ \bibinfo {author} {\bibfnamefont {J.}~\bibnamefont {Cao}},\
  }\href@noop {} {\bibfield  {journal} {\bibinfo  {journal} {J. Chem. Phys.}\
  }\textbf {\bibinfo {volume} {139}},\ \bibinfo {pages} {134106} (\bibinfo
  {year} {2013})}\BibitemShut {NoStop}%
\bibitem [{\citenamefont {Tang}\ \emph {et~al.}(2015)\citenamefont {Tang},
  \citenamefont {Ouyang}, \citenamefont {Gong}, \citenamefont {Wang},\ and\
  \citenamefont {Wu}}]{Tang2015Extendedhierarchyequation}%
  \BibitemOpen
  \bibfield  {author} {\bibinfo {author} {\bibfnamefont {Z.}~\bibnamefont
  {Tang}}, \bibinfo {author} {\bibfnamefont {X.}~\bibnamefont {Ouyang}},
  \bibinfo {author} {\bibfnamefont {Z.}~\bibnamefont {Gong}}, \bibinfo {author}
  {\bibfnamefont {H.}~\bibnamefont {Wang}}, \ and\ \bibinfo {author}
  {\bibfnamefont {J.}~\bibnamefont {Wu}},\ }\href
  {http://scitation.aip.org/content/aip/journal/jcp/143/22/10.1063/1.4936924}
  {\bibfield  {journal} {\bibinfo  {journal} {J. Chem. Phys.}\ }\textbf
  {\bibinfo {volume} {143}},\ \bibinfo {pages} {224112} (\bibinfo {year}
  {2015})}\BibitemShut {NoStop}%
\bibitem [{\citenamefont {Ye}\ \emph {et~al.}(2016)\citenamefont {Ye},
  \citenamefont {Wang}, \citenamefont {Hou}, \citenamefont {Xu}, \citenamefont
  {Zheng},\ and\ \citenamefont {Yan}}]{Ye2016HEOMQUICKprogram}%
  \BibitemOpen
  \bibfield  {author} {\bibinfo {author} {\bibfnamefont {L.}~\bibnamefont
  {Ye}}, \bibinfo {author} {\bibfnamefont {X.}~\bibnamefont {Wang}}, \bibinfo
  {author} {\bibfnamefont {D.}~\bibnamefont {Hou}}, \bibinfo {author}
  {\bibfnamefont {R.-X.}\ \bibnamefont {Xu}}, \bibinfo {author} {\bibfnamefont
  {X.}~\bibnamefont {Zheng}}, \ and\ \bibinfo {author} {\bibfnamefont
  {Y.}~\bibnamefont {Yan}},\ }\href {\doibase 10.1002/wcms.1269} {\bibfield
  {journal} {\bibinfo  {journal} {Wiley Interdiscip. Rev.: Comput. Mol. Sci.}\
  }\textbf {\bibinfo {volume} {6}},\ \bibinfo {pages} {608} (\bibinfo {year}
  {2016})}\BibitemShut {NoStop}%
\bibitem [{\citenamefont {H{\"a}rtle}\ \emph {et~al.}(2015)\citenamefont
  {H{\"a}rtle}, \citenamefont {Cohen}, \citenamefont {Reichman},\ and\
  \citenamefont {Millis}}]{Haertle2015TransportthroughAnderson}%
  \BibitemOpen
  \bibfield  {author} {\bibinfo {author} {\bibfnamefont {R.}~\bibnamefont
  {H{\"a}rtle}}, \bibinfo {author} {\bibfnamefont {G.}~\bibnamefont {Cohen}},
  \bibinfo {author} {\bibfnamefont {D.~R.}\ \bibnamefont {Reichman}}, \ and\
  \bibinfo {author} {\bibfnamefont {A.~J.}\ \bibnamefont {Millis}},\ }\href
  {\doibase 10.1103/PhysRevB.92.085430} {\bibfield  {journal} {\bibinfo
  {journal} {Physical Review B}\ }\textbf {\bibinfo {volume} {92}},\ \bibinfo
  {pages} {085430} (\bibinfo {year} {2015})}\BibitemShut {NoStop}%
\bibitem [{\citenamefont {Segal}, \citenamefont {Millis},\ and\ \citenamefont
  {Reichman}(2010)}]{Segal2010Numericallyexactpath}%
  \BibitemOpen
  \bibfield  {author} {\bibinfo {author} {\bibfnamefont {D.}~\bibnamefont
  {Segal}}, \bibinfo {author} {\bibfnamefont {A.~J.}\ \bibnamefont {Millis}}, \
  and\ \bibinfo {author} {\bibfnamefont {D.~R.}\ \bibnamefont {Reichman}},\
  }\href {\doibase 10.1103/PhysRevB.82.205323} {\bibfield  {journal} {\bibinfo
  {journal} {Phys. Rev. B}\ }\textbf {\bibinfo {volume} {82}},\ \bibinfo
  {pages} {205323} (\bibinfo {year} {2010})}\BibitemShut {NoStop}%
\bibitem [{\citenamefont {Wang}\ and\ \citenamefont
  {Thoss}(2013)}]{Wang2013Numericallyexacttime}%
  \BibitemOpen
  \bibfield  {author} {\bibinfo {author} {\bibfnamefont {H.}~\bibnamefont
  {Wang}}\ and\ \bibinfo {author} {\bibfnamefont {M.}~\bibnamefont {Thoss}},\
  }\href
  {http://scitation.aip.org/content/aip/journal/jcp/138/13/10.1063/1.4798404}
  {\bibfield  {journal} {\bibinfo  {journal} {J. Chem. Phys.}\ }\textbf
  {\bibinfo {volume} {138}},\ \bibinfo {pages} {134704} (\bibinfo {year}
  {2013})}\BibitemShut {NoStop}%
\bibitem [{\citenamefont {Gull}, \citenamefont {Reichman},\ and\ \citenamefont
  {Millis}(2010)}]{Gull2010Boldlinediagrammatic}%
  \BibitemOpen
  \bibfield  {author} {\bibinfo {author} {\bibfnamefont {E.}~\bibnamefont
  {Gull}}, \bibinfo {author} {\bibfnamefont {D.~R.}\ \bibnamefont {Reichman}},
  \ and\ \bibinfo {author} {\bibfnamefont {A.~J.}\ \bibnamefont {Millis}},\
  }\href {\doibase 10.1103/PhysRevB.82.075109} {\bibfield  {journal} {\bibinfo
  {journal} {Phys. Rev. B}\ }\textbf {\bibinfo {volume} {82}},\ \bibinfo
  {pages} {075109} (\bibinfo {year} {2010})}\BibitemShut {NoStop}%
\bibitem [{\citenamefont {Segal}\ and\ \citenamefont
  {Nitzan}(2005)}]{Segal2005SpinBosonThermal}%
  \BibitemOpen
  \bibfield  {author} {\bibinfo {author} {\bibfnamefont {D.}~\bibnamefont
  {Segal}}\ and\ \bibinfo {author} {\bibfnamefont {A.}~\bibnamefont {Nitzan}},\
  }\href {\doibase 10.1103/PhysRevLett.94.034301} {\bibfield  {journal}
  {\bibinfo  {journal} {Phys. Rev. Lett.}\ }\textbf {\bibinfo {volume} {94}},\
  \bibinfo {pages} {034301} (\bibinfo {year} {2005})}\BibitemShut {NoStop}%
\bibitem [{\citenamefont {Nicolin}\ and\ \citenamefont
  {Segal}(2011)}]{Nicolin2011Nonequilibriumspin}%
  \BibitemOpen
  \bibfield  {author} {\bibinfo {author} {\bibfnamefont {L.}~\bibnamefont
  {Nicolin}}\ and\ \bibinfo {author} {\bibfnamefont {D.}~\bibnamefont
  {Segal}},\ }\href {\doibase 10.1063/1.3655674} {\bibfield  {journal}
  {\bibinfo  {journal} {J. Chem. Phys.}\ }\textbf {\bibinfo {volume} {135}},\
  \bibinfo {pages} {164106} (\bibinfo {year} {2011})}\BibitemShut {NoStop}%
\bibitem [{\citenamefont {Saito}\ and\ \citenamefont
  {Kato}(2013)}]{Saito2013KondoSignaturein}%
  \BibitemOpen
  \bibfield  {author} {\bibinfo {author} {\bibfnamefont {K.}~\bibnamefont
  {Saito}}\ and\ \bibinfo {author} {\bibfnamefont {T.}~\bibnamefont {Kato}},\
  }\href {\doibase 10.1103/PhysRevLett.111.214301} {\bibfield  {journal}
  {\bibinfo  {journal} {Phys. Rev. Lett.}\ }\textbf {\bibinfo {volume} {111}},\
  \bibinfo {pages} {214301} (\bibinfo {year} {2013})}\BibitemShut {NoStop}%
\bibitem [{\citenamefont {Velizhanin}, \citenamefont {Wang},\ and\
  \citenamefont {Thoss}(2008)}]{Velizhanin2008Heattransportthrough}%
  \BibitemOpen
  \bibfield  {author} {\bibinfo {author} {\bibfnamefont {K.~A.}\ \bibnamefont
  {Velizhanin}}, \bibinfo {author} {\bibfnamefont {H.}~\bibnamefont {Wang}}, \
  and\ \bibinfo {author} {\bibfnamefont {M.}~\bibnamefont {Thoss}},\ }\href
  {\doibase 10.1016/j.cplett.2008.05.065} {\bibfield  {journal} {\bibinfo
  {journal} {Chem. Phys. Lett.}\ }\textbf {\bibinfo {volume} {460}},\ \bibinfo
  {pages} {325} (\bibinfo {year} {2008})}\BibitemShut {NoStop}%
\bibitem [{\citenamefont {Velizhanin}, \citenamefont {Thoss},\ and\
  \citenamefont {Wang}(2010)}]{Velizhanin2010MeirWingreenformulaheat}%
  \BibitemOpen
  \bibfield  {author} {\bibinfo {author} {\bibfnamefont {K.~A.}\ \bibnamefont
  {Velizhanin}}, \bibinfo {author} {\bibfnamefont {M.}~\bibnamefont {Thoss}}, \
  and\ \bibinfo {author} {\bibfnamefont {H.}~\bibnamefont {Wang}},\ }\href
  {\doibase 10.1063/1.3483127} {\bibfield  {journal} {\bibinfo  {journal} {J.
  Chem. Phys.}\ }\textbf {\bibinfo {volume} {133}},\ \bibinfo {pages} {084503}
  (\bibinfo {year} {2010})}\BibitemShut {NoStop}%
\end{thebibliography}%

\end{document}